\documentclass[a4paper,11pt]{article}
\pdfoutput=1
\usepackage{jcappub}
\usepackage{graphicx}
\usepackage{subcaption}
\usepackage{gensymb}
\usepackage{multirow}
\usepackage{float}
\usepackage{mathtools}
\usepackage{comment,cancel}
\usepackage{indentfirst}
\usepackage{enumerate}

\usepackage[T1]{fontenc}
\usepackage{booktabs}
\usepackage[usenames,dvipsnames,table]{xcolor}
\usepackage{xfrac}
\usepackage{placeins}

\newcommand\dd{\textrm{d}}

\newcommand{\hunit}{{\rm km}\,{\rm s}^{-1}\,{\rm Mpc}^{-1}}

\title{A robust cosmic standard ruler from the cross-correlations of galaxies and dark sirens}

\author[a]{Jo\~ao Ferri}
\author[a]{, Ian L. Tashiro}
\author[a]{, L. Raul Abramo}
\author[b,c,g]{, Isabela Matos}
\author[d,e,f]{, Miguel Quartin}
\author[b,d,g]{, Riccardo Sturani}

\affiliation[a]{\small Departamento de F\'{\i}sica Matem\'atica, Instituto de F\'{\i}sica, Universidade de S\~ao Paulo,\\ R.  do  Mat\~ao  1371,  05508-090,  S\~ao Paulo, SP, Brazil}

\affiliation[b]{\small Instituto de Física Teórica, Universidade Estadual Paulista, São Paulo 01140-070, SP, Brazil.}

\affiliation[c]{\small Institute of Cosmology and Gravitation, University of Portsmouth,
Burnaby Road, Portsmouth PO1 3FX, UK}

\affiliation[d]{\small PPG Cosmo, Universidade Federal do Espírito Santo, 29075-910, Vitória, ES, Brazil.}

\affiliation[e]{\small Instituto de Física, Universidade Federal do Rio de Janeiro, 21941-972, Rio de Janeiro, RJ, Brazil.}

\affiliation[f]{\small Observatório do Valongo, Universidade Federal do Rio de Janeiro, 20080-090, Rio de Janeiro, RJ, Brazil.}

\affiliation[g]{\small ICTP South American Institute for Fundamental Research, São Paulo 01140-070, SP, Brazil.}

\emailAdd{joao.vitor.ferri@usp.br}
\emailAdd{iantashiro@usp.br}
\emailAdd{raulabramo@usp.br}
\emailAdd{isabela.matos@port.ac.uk}
\emailAdd{mquartin@if.ufrj.br}
\emailAdd{riccardo.sturani@unesp.br}

\abstract{
Observations of gravitational waves (GWs) from dark sirens allow us to infer their locations and distances.
Galaxies, on the other hand, have precise angular positions but no direct measurement of their distances -- only redshifts.
The cross-correlation of GWs, which we limit here to binary black hole mergers (BBH), in spherical shells of luminosity distance $D_L$, with galaxies in shells of redshift $z$, leads to a direct measurement of the Hubble diagram $D_L(z)$.
Since this standard ruler relies only on the statistical proximity of the dark sirens and galaxies (a general property of large-scale structures), it is essentially model-independent: the correlation is maximal when both redshift and $D_L$ shells coincide.
We forecast the constraining power of this technique, which we call {\it{Peak Sirens}},  for run~5~(O5) of LIGO-Virgo-KAGRA (LVK), as well as for the third-generation observatories Einstein Telescope and Cosmic Explorer.
We employ thousands of full-sky light cone simulations with realistic numbers for the tracers, and include masking by the Milky Way, lensing and inhomogeneous GW sky coverage.
We find that the method is not expected to suffer from some of the issues present in other dark siren methods, such as biased constraints due to incompleteness of galaxy catalogs or dependence on priors for the merger rates of BBH.
We show that with Peak Sirens, given the projected O5 sensitivity, LVK can measure $H_0$ with $7\%$ precision by itself, assuming $\Lambda$CDM, and $4\%$ precision using external datasets to constrain $\Omega_m$.
We also show that future third-generation GW detectors can achieve, without external data, sub-percent uncertainties in $H_0$ assuming $\Lambda$CDM, and 3\% in a more flexible $w_0w_a$CDM model. The method also shows remarkable robustness against systematic effects such as the modeling of non-linear structure formation.
}

\keywords{Large Scale Structure, Angular Power Spectrum, Multi-tracer, Dark Sirens}

\makeatletter
\gdef\@fpheader{}
\makeatother

\begin{document}

%\preprint{
\begin{flushright}
{\small{ET-0544A-24} }
\end{flushright}

\maketitle
\flushbottom

%%%%%%%%%%%%%%%%%%%%%%%%%%%%%
\section{Introduction}
\label{sec:introduction}

The large-scale structures of the Universe are increasingly being mapped with multiple tracers and messengers, and the complementarity between different surveys has enormous potential to test cosmological models and resolve some of the existing tensions of the standard $\Lambda$CDM scenario~\cite{H0Review3,Sigma8}.

\begin{figure}[t!]
    \centering
    \includegraphics[width=0.42\columnwidth]{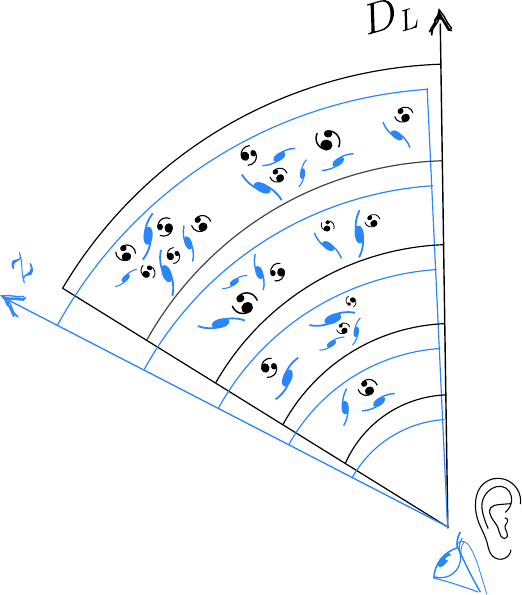} $\;\;$
    \includegraphics[width=0.54\columnwidth]{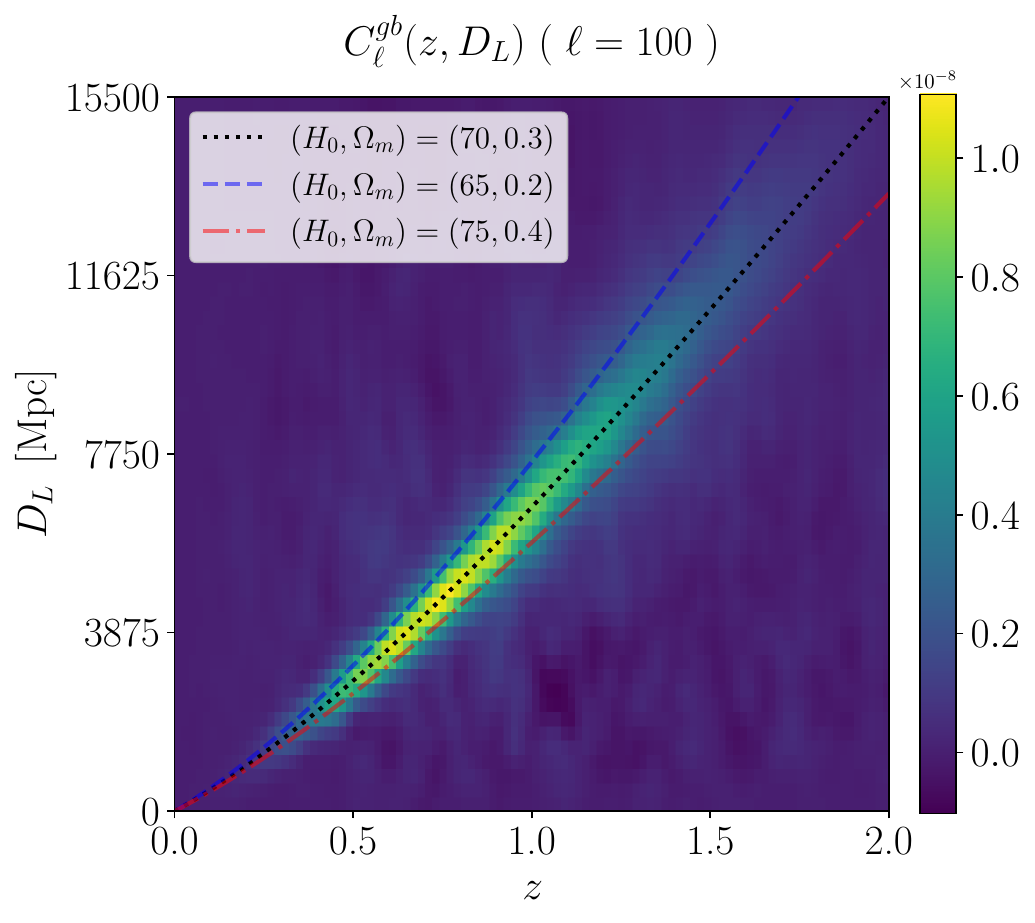}
    \caption{\emph{Left:} A galaxy map (in redshift $z$) overlaps a BBH map (in luminosity distance $D_L$). When a shell in $z$ coincides with a shell in $D_L$, the correlation between both maps is maximal. \emph{Right:} The cross-correlation between the harmonic modes of galaxies (index $g$) at $z$ and BBHs (index $b$) at $D_L$ leads to a cross-angular power spectrum $C^{gb}_\ell(z,D_L)$.
    We show the average cross-correlation for 1000 light-cone simulations of galaxies and BBH maps, for $\ell=100$ ($\sim 2^\circ$).
    The red, blue and black lines corresponding to the Hubble diagrams $D_L(z)$ of three different cosmologies -- in our fiducial model $H_0=70.0\,\, {\rm km}\, {\rm s}^{-1}\, {\rm Mpc}^{-1}$ and $\Omega_m=0.3$.      \label{fig:Intro12}}
\end{figure}

In this paper we investigate
how we can exploit the cross-correlations between redshift surveys and detections of gravitational waves (GW) from dark sirens (most of which are binary black hole mergers, BBH) to trace the Hubble distance diagram $D_L(z)$ -- even when the galaxies that host GWs are absent from the redshift survey.
The idea is illustrated in Fig.~\ref{fig:Intro12}: a galaxy map, where the radial information is only in the redshifts, covers the same area in the sky as a map of GWs, for which the gravitational wave (luminosity) distances can be inferred from the waveforms.

Without electromagnetic counterparts, which only occur for the very rare bright standard sirens, direct redshift measurements are in general not possible for the dark sirens.
In the absence of a cosmological model, the maps of galaxies and of dark sirens are disjoint: the relationship between redshifts and distances is only determined once the cosmology is fixed.
It is useful to regard the problem in terms of spherical shells (or bins), noting that we do not know {\em a priori} how the spherical shells in redshift are mapped onto the shells in the gravitational wave (luminosity) distances.

However, galaxies and BBHs are connected by the underlying large-scale structures of the Universe, as manifested by their two-point correlation functions.
In fact, even if all the galaxies that host BBHs are absent from the galaxy catalog, there would still be a correlation between the two surveys due to the surrounding structures \cite{bera2020incompleteness}. Since the clustering (the amplitude of the matter correlation function) is larger for small distances, the angular cross-correlation between galaxies (in shells of redshift) and BBHs (in shells of distance) peaks when the redshift shells {\it match} the distance shells -- i.e., when we have the true underlying cosmology determining the relationship $D_L(z)$.
The detection of this peak constitutes a cosmic ruler, which then leads to constraints on $H_0$ and other cosmological parameters \cite{oguri2016measuring,bera2020incompleteness,Ghosh:2023ksl}.

The novel way in which we have implemented this method, which we call {\it{Peak Sirens}}, is shown in Fig.~\ref{fig:Intro12}, where we plot the angular power spectrum for galaxies (in bins of redshift) and BBHs (in bins of distance), obtained from an average over 1000 full-sky simulations of our past light cone. It is clear that the peak of the correlation is precisely at the locus of the Hubble diagram for the fiducial model ($H_0=70.0\,\, {\rm km}\, {\rm s}^{-1}\, {\rm Mpc}^{-1}$ and $\Omega_m=0.3$). As we show in this paper, the position of the peak of the cross-correlation traces the Hubble diagram and is the universal feature that determines the underlying cosmological background.

Cross-correlating BBHs with redshift surveys is one of several methods for extracting cosmological information from standard siren observations. Notably, the most direct way of measuring cosmological parameters or testing the cosmological model with GWs relies on the detection of bright standard sirens, as demonstrated by the impact of the optical transient AT 2017gfo associated with the event GW170817 \cite{LIGOScientific:2017vwq, LIGOScientific:2017ync, LIGOScientific:2018dkp} and its other counterparts in various bands. However, there are several challenges to the bright siren approach. For the current observing facilities, the BNS horizon is still very shallow, which results in the current ratio of dark to bright sirens of over $100$. Second, for next generation GW observatories, the observed BNS events at high redshift should have very faint electromagnetic counterparts, which means that the number of bright sirens will be limited by the available telescope infrastructure for follow-up surveys~\cite{Belgacem:2019tbw,Chen:2020zoq,Alfradique:2022tox,2022MNRAS.511.2782C,Afroz:2024joi}.
This means that dark sirens could potentially be the main target for standard siren cosmology and have already been analyzed with available observations \cite{GWTC3_constraints,LIGOScientific:2020zkf}.
So far, the most studied dark siren methods are the so-called spectral sirens, which take redshift information from features in the detector frame mass distribution of black-holes such as mass gaps \cite{Taylor:2011fs, Wysocki:2018mpo, Farr:2019twy, Mastrogiovanni:2021wsd, Mancarella:2021ecn, Mukherjee:2021rtw,Ezquiaga:2022zkx,Karathanasis:2022rtr}, and the line-of-sight prior method \cite{2012PhRvD..86d3011D, LIGOScientific:2018gmd, PhysRevD.101.122001}, which draws from the same basic principle of the cross-correlation methods -- i.e., the fact that GW sources are expected to be located inside galaxies.

The line-of-sight method relies on the use of galaxy catalogs from which one constructs priors on the redshifts of GW sources, by weighting the redshifts of all potential host galaxies within the GW detection volume by some modeled probability of them being the true host of the GW source (see \cite{2023AJ....166...22G} for a pedagogical review, or \cite{OShaughnessy:2009szr,Leandro:2021qlc,Hanselman:2024hqy,Perna:2024lod,Vijaykumar:2023bgs,Mapelli:2019bnp,Artale:2019doq} for studies about the impact of redshift evolution of galaxy properties). Apart from the computational challenge of including the many galaxies within the large areas defined by the GW posteriors, incompleteness of galaxies catalogs at relatively low redshifts has been the main difficulty for this method \cite{Finke:2021aom}.
To work this out consistently, the probability of the BBHs not belonging to any of the galaxies in the catalog must be included in the likelihood, in which case one needs to add a prior in their redshifts motivated by population models for black holes.
These unknown properties, notably the BBH merger rate, have then to be measured jointly with cosmological parameters, due to significant correlations \cite{Gray:2023wgj,2023PhRvD.108d2002M}.

Cross-correlating galaxies in redshift space with other tracers of known (luminosity) distances was originally proposed as a cosmological test by Ref. \cite{oguri2016measuring}.
In Ref. ~\cite{MukherjeeWandelt2018} the authors computed Fisher forecasts for the potential of cross-correlating Type 1a supernovas with galaxies to constrain dark energy parameters, and  Refs. ~\cite{Mukherjee:2019wcg,Mukherjee:2020hyn} study cosmological constraints from cross-correlations of GWs with redshift tracers. However, while we take into account the correlations between objects at {\em different} radial positions, and locate the positions where those correlations peak in the $(z,D_L)$ plane, those previous works explored instead the {\it amplitude} of the correlations between objects at the same redshifts, fitting the cosmology through the matter power spectrum.
Recently, Ref.~\cite{Mukherjee:2022afz} studied how dark sirens from the GWTC-3 \cite{2023PhRvX..13d1039A} can be combined with galaxies to constrain $H_0$ and other cosmological parameters, deriving constraints from each individual GW event by using the assumed form of the cross-correlations as priors to the redshifts of those events -- for a related approach, see also Ref. \cite{bera2020incompleteness}.

As we show in this paper, the angular cross-correlation technique that underlies the Peak Sirens method is basically model-independent -- that is, matching the two large-scale structures in different spaces depends only mildly on the details of the individual structures themselves.
To demonstrate the full potential of Peak Sirens as a cosmic standard ruler we have introduced a novel approach which includes: (i) full-sky simulations of the light cone with both galaxies and dark sirens, including gravitational lensing, possible thanks to the latest version of the \texttt{GLASS} code\footnote{\url{https://github.com/glass-dev/glass}} \cite{tessore2023glass};
(ii) the numbers and spatial distribution of dark sirens, as well as the uncertainties in their angular and radial positions, are those predicted for second and third-generation experiments by state-of-the-art Fisher forecasts for GW parameter estimation;
(iii) a partial sky coverage due to the mask imposed by the Milky Way on the galaxy redshift surveys;
(iv) thin redshift and distance slices to better capture the radial information from galaxies and black holes, including the cross-correlations between different slices;
and (v) an accelerated method for computing the angular power spectra in different cosmological scenarios, which allows us to explore a wide parameter space and a fuller range of physical scales within a Markov Chain Monte Carlo (MCMC) statistical inference approach.

We have deployed this machinery to study future constraints in three scenarios. From the most current to the most futuristic, they are: the fifth observation run (O5) of LIGO-Virgo-KAGRA (LVK) \cite{LIGOScientific:2014pky,VIRGO:2014yos,KAGRA:2020tym}; the current baseline configuration for the third-generation network composed by the triangular Einstein Telescope (ET) combined with LVK at its design sensitivity, running for 5 years (LVK+ET); and ET combined with two Cosmic Explorer (CE) detectors, also in a 5-year run (ET+2CE)~\cite{Maggiore:2019uih,Reitze:2019iox,Kalogera:2021bya}.

The structure of the paper is as follows. In section \ref{sec:AngCorrGeneral} we describe the formalism for multi-tracer analysis and the core of the method employed in this paper. In section \ref{sec:simulations} we detail the various ingredients included in the construction of both galaxy and BBH simulations. In section \ref{sec:results} we present our results and in section \ref{sec:conclusion} we conclude. We further discuss technical aspects of the simulations in the Appendix \ref{sec:appendix}.

%%%%%%%%%%%%%%%%%%%%%%%%%%%%%
\section{The angular correlation function and angular power spectrum}
\label{sec:AngCorrGeneral}
%%%%%%%%%%%%%%%%%%%%%%%%%%%%%

Consider a tracer of large-scale structure at some radial bin $r_i$ -- in the case of galaxies, $r_i$ would be a redshift bin (or slice), while for BBHs $r_i$ correspond to a bin in luminosity distance.
The counts of a tracer $\alpha$ (red galaxies, emission line galaxies, BBHs, etc.) can be denoted by $N^{\alpha}(r_i,\vec\Omega_i)$ where $\vec\Omega$ is the angular position $(\theta,\phi)$ of each object.
We then define the contrast map as:
\begin{equation}
\label{eqn:ContrastMap}
    \Delta N^{\alpha}(r_i,\vec\Omega_i)=N^{\alpha}(r_i,\vec\Omega_i)-\bar{N}^\alpha(r_i) \, .
\end{equation}
where $\bar{N}^\alpha(r_i)$ is the average over the angular positions (we assume here for simplicity that the angular selection function is independent of the position in the sky).

For mergers of compact objects observed through their gravitational waves, whose distances are known but redshifts are not, those radial labels are the luminosity distance, $r \to D_L$. Here we assume that the  distance that can be inferred from GW observations is the luminosity distance, which is true in General Relativity but may change in modified gravity models~\cite{Lombriser:2015sxa, Amendola:2017ovw, Belgacem:2018lbp, Nishizawa:2017nef}. This is in fact the main deviation from the standard cosmological model that can be probed by GW detectors in a model-independent way (see e.g.~\cite{Matos:2023jkn}), which could also in principle be done with the cross-correlation method. Although currently the overwhelming majority of GW detections are classified as BBHs, this is likely due to a selection bias of the detectors~\cite{LIGOScientific:2021psn}, which are at present only sensitive to binary neutron star mergers (BNS)  for $z < 0.04$, and slightly deeper for neutron star black hole mergers (NSBH). Here we consider only BBHs since they are expected to be more abundant, and to cover a deeper redshift range, but we note that our proposed method works irrespective of the GW source.

The contrast map can then be expressed in terms of spherical harmonics:
\begin{equation}
    \label{eqn:ContrastDecomp}
    \Delta N^{\alpha}(r_i,\vec\Omega_i)=\sum_{\ell=1}^{\infty}\sum_{m=-\ell}^{\ell}\,
    a^{\alpha}_{\ell\,m}(r_i)\,Y_{\ell\,m}\,(\vec\Omega_i) \; .
\end{equation}
The correlation of these harmonic modes between two tracers $\alpha$ and $\beta$ is then given by:
\begin{equation}
    \label{eqn:CrossCorr}
    \langle a^{\alpha}_{\ell\,m}(r_i)\,a^{\beta*}_{\ell'\,m'}(r_j)\rangle=\delta_{\ell\,\ell'}\,\delta_{m\,m'}\,\Gamma_\ell^{\alpha \beta} (r_i,r_j) \, ,
\end{equation}
where the statistical mean above is in practice approximated by the average over $m$ modes for the correlation coefficients:
\begin{equation}
    \label{eqn:cros_corr}
    \Gamma_\ell^{\alpha \beta} (r_i,r_j) = \frac{1}{2\ell+1}
    \sum_{m=-\ell}^\ell
    a^{\alpha}_{\ell\,m}(r_i)\, a^{\beta \, *}_{\ell\,m}(r_j)\, .
\end{equation}
We obtain the angular power spectrum after subtracting the shot noise (assumed Poissonian here) from the correlation coefficients above:
\begin{equation}
    \label{eqn:angpow}
    C_\ell^{\alpha \beta} (r_i,r_j)=\Gamma_\ell^{\alpha\beta} (r_i,r_j)-\delta_{ij}\delta^{\alpha\beta}\,\bar{N}^{\alpha}(r_i) \; .
\end{equation}
The correlation function in configuration space can be written in terms of the angular power spectrum by:
\begin{equation}
    \label{eqn:RealSpaceCorr}
    \xi^{\alpha\beta}(\Vec{r}_i,\Vec{r}_j)=\sum_\ell\,\frac{2\ell+1}{4\pi}\, C_\ell^{\alpha\beta} (r_i,r_j)\,P_\ell(\hat{r}_i\cdot\hat{r}_j)\,,
\end{equation}
\begin{comment}
\begin{equation}
    \label{eqn:RealSpaceCorr}
    \xi^{ij}(\Vec{r}_i,\Vec{r}_j)=\sum_\ell\,\frac{2\ell+1}{4\pi}\, C_\ell^{ij} (r_i,r_j)\,P_\ell(\hat{r}_i\cdot\hat{r}_j)\,,
\end{equation}
\end{comment}
where $P_\ell$ are the Legendre Polynomials.

\begin{figure}
    \centering
    \includegraphics[width=0.67\linewidth,trim={0cm 0.9cm 0cm 0.9cm}, clip]{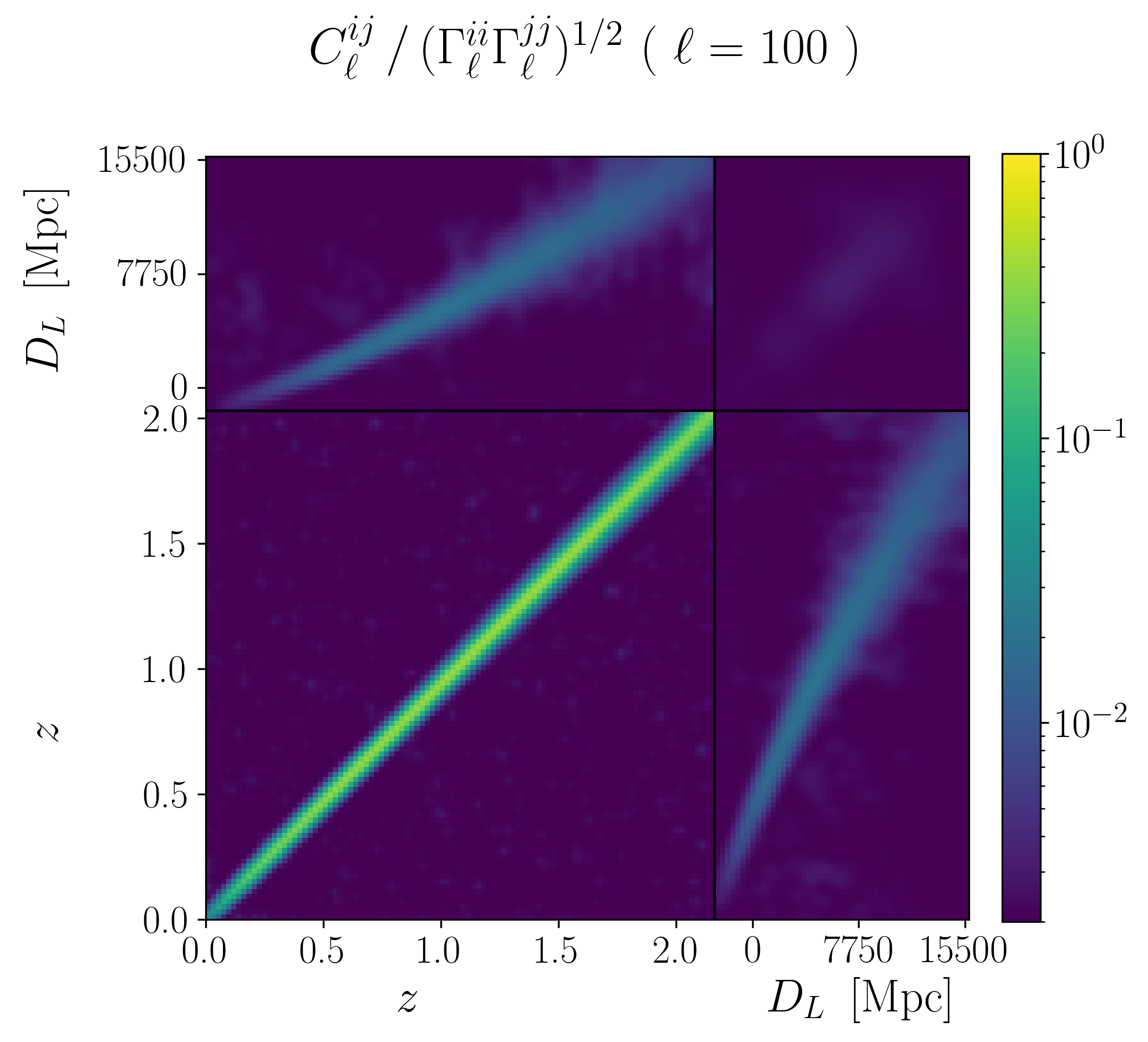}
    \caption{Multi-tracer angular power spectrum $C_\ell^{ij}$ for $\ell=100$, normalized by $\sqrt{\Gamma^{ii}\Gamma^{jj}}$ in order to highlight the signal-to-noise ratio (using averages over 1000 simulations). Following~\eqref{eqn:radialbins}, $C_\ell^{ij}$ can be decomposed into four blocks. Our observable of interest, the galaxy-BBH cross-correlations $C_\ell^{gb}(z,D_L)$, is contained in the off-diagonal blocks. The diagonal blocks are the auto-correlations for galaxy-galaxy $C_\ell^{gg}(z,z)$ (lower left) and BBH-BBH $C_\ell^{bb}(D_L,D_L)$ (upper right, dominated by shot-noise), both of which we disregard in our analysis.}
    \label{fig:et2ce_corrl}
\end{figure}

Notice that the indices for tracers ($\alpha,\beta$) and slices ($i,j$) are in fact redundant: a single index can refer to both a radial slice and a tracer in that slice.
Therefore, we simplify our notation by dropping the tracer indices $\alpha$ and $\beta$, henceforth assuming that whenever a tracer is in a redshift bin it corresponds to the galaxies, and if it is in a distance bin, it corresponds to BBHs.
For clarity, given $N_z$ redshift bins and $N_d$ distance bins, we define $N_r \equiv N_z + N_d$ radial bins in the following way:
\begin{equation}
    \label{eqn:radialbins}
    \{ r \} = \{ z_1 , z_2, \ldots, z_{N_z} , D_1 , D_2, \ldots D_{N_d} \} \; .
    \end{equation}
In this way we can write without any ambiguity $C_\ell^{\alpha\beta} (r_i,r_j) \to C_\ell^{ij}$ (see Fig.~\ref{fig:et2ce_corrl}).
In fact, this notation can be generalized for an arbitrary number of tracers, and any number of radial bins for each tracer: with $T$ tracers, each one sliced into $N_t$ radial bins, the entire set of radial bins is an array of length $N_r = \sum_{t=1}^T N_t$.
It is important to stress that each tracer can be binned into arbitrary redshift or distance slices -- in particular, it is irrelevant whether or not the redshift and the distance bins coincide (in general they do not, since we do not know {\em a priori} the cosmological model).

\subsection{Multi-tracer likelihoods}

Here, we employ the angular cross spectrum of galaxies and BBHs, $C_\ell^{ij}$, as our physical observable, and derive constraints based on its expected (theoretical) values. In terms of a set of cosmological parameters $\theta$, the logarithm of the likelihood is expressed as $-2 \log {\cal{L}} \to \chi^2(\theta)$:
\begin{equation}
    \label{eqn:MultiTracerChi2}
    \chi^2(\theta)= \frac{f_{\rm sky}}{2}
    \sum_{i j} \sum_{i' j'}
    \, \sum_{\ell,\ell'}\,
    \Delta\,C_\ell^{ij}
    \,{\rm Cov}^{-1} [C_\ell^{ij},C_{\ell'}^{i'j'}] \,
    \Delta\,C_\ell^{i'j'}\,,
\end{equation}
where
\begin{equation}
    \label{eqn:Delta_Cl}
    \Delta\,C_\ell^{ij} \equiv C_\ell^{ij,{\rm obs}}-C_\ell^{ij} (\theta)\, ,
\end{equation}
$f_{\rm sky}$ is the observed fraction of the sky, and the factor of $1/2$ takes into account the fact that $C_\ell^{ij}$ is a symmetric matrix, whose  degrees of freedom must be counted only one time.

Computing the covariance of such a high-dimensional object can be challenging, as angular and radial modes generally mix in a non-trivial way. If the data uniformly covers the entire sky, then each angular mode $\ell$ is independent of the others, leading to a covariance structure where ${\rm Cov}[C_\ell^{ij}, C_{\ell'}^{i'j'}] \propto \delta_{\ell \ell'}$. However, the presence of an angular mask disrupts this orthogonality. In our approach, we work with bins in $\ell$ that minimize the covariance between bins for the specific mask we employ (a simple cut in galactic latitude) -- for more details, see the Appendix. Regarding the radial dependence, when radial bins are small enough (as is the case in our method) correlations between nearby slices arise and must be included in the covariance. In particular, for a fixed $\ell$ and $N_r$ radial bins (where the radial binning also determines the tracer), each angular power spectrum $C_\ell^{ij}$ has $N_r(N_r+1)/2$ degrees of freedom
(with a mask the different $\ell$ modes become correlated, in which case the number of degrees of freedom can be smaller -- see the Appendix).

It is instructive to estimate the number of degrees of freedom for the covariance of the angular power spectra. Assuming that observations allow us to resolve angular scales up to some $\ell_{\rm max}$, the covariance has $\sim~\ell_{\rm max}\times (N_r^{2}/2)^2$ non-trivial entries. If we limit ourselves to relatively large scales, $\ell_{\rm max}=100$, using  $N_{r} \sim 150$ radial bins leads to $\sim 10^{10}$ degrees of freedom for the covariance.  Even if there is significant redundancy in the covariance, we would still need millions of simulations for a numerical (sample) covariance matrix to converge, which is not feasible even considering simplified mock simulations.

The alternative is to employ analytical or semi-analytical  approximations for the covariance. Here we take a hybrid approach: first, we simulate light cones with galaxies and BBHs to compute a sample mean of $\Gamma_\ell^{ij}$ (the angular power spectrum including shot noise). Then, we use the Gaussian approximation to compute ${\rm Cov}[C_\ell^{ij},C_\ell^{i'j'}]$ and its inverse. In this framework, the Fisher matrix ${\rm Cov}[C_\ell^{ij},C_\ell^{i'j'}]^{-1}\equiv\,F_\ell[C_\ell^{ij},C_\ell^{i'j'}]$ is given by \cite{AFT2022a,AFT2022b}:
\begin{eqnarray}
\label{Eq:FisherMatrix}
  F[C^{ij}_\ell,C^{i'j'}_\ell]
  &=&
  \frac{2\ell+1}{4}
  (2-\delta_{ij}) (2-\delta_{i'j'})
   \left\{
   [{\Gamma}_{\ell}^{-1}]^{i i'}
   [{\Gamma}_{\ell}^{-1}]^{j j'}
   +
   [{\Gamma}_{\ell}^{-1}]^{i j'}
   [{\Gamma}_{\ell}^{-1}]^{i' j}
  \right\}   .
\end{eqnarray}

Notice that \eqref{Eq:FisherMatrix} includes information from both the auto- and the cross-correlations.
Since we are interested in quantifying the constraining power due to cross-correlations alone,
we take two important steps: (i) we marginalize over the signal of the auto-correlations by keeping only the noise in both $\Gamma_\ell^{z_g \, z_g'} \to \Gamma_\ell^{{\rm N}, \, z_g \, z_g'}$ and $\Gamma_\ell^{D_L \, D_L'} \to \Gamma_\ell^{{\rm N}, \, D_L \, D_L'}$ before inverting the matrices $\mathbf{\Gamma}^{-1}$ in Eq.~\eqref{Eq:FisherMatrix}, where $z_g$ and $D_L$ are, respectively, the bins in redshift (of galaxies) and in distance (of BBHs); and (ii) we will consider as our data only the cross-correlations $\mathbf{C}^{\rm x}_\ell = \{ C_\ell^{z_g \, D_L} \, , \, C_\ell^{D_L \, z_g} \}$.
Applying \eqref{Eq:FisherMatrix} into \eqref{eqn:MultiTracerChi2}, marginalizing over the signal of the auto-correlations and limiting the data vector to the cross-correlations, we obtain:
\begin{eqnarray}
    \label{eq:gal_bbh_chi2_matrix}
    \chi^2(\theta) &=& \,
    \frac{f_{\rm sky}}{2}  \sum_{\ell} \,
    b_{\ell}\, (2\ell+1)\,
    \sum_{ii' \in z_g} \sum_{jj' \in D_L} \\
    \nonumber
    & {\,} & \times
    \left[
    \Delta C_\ell^{ij}
    \left(\Gamma^{{\rm N}, \, -1}_\ell\right)^{j j'}
    \Delta C_\ell^{j'i'}
    \left(\Gamma^{{\rm N}, \, -1}_\ell\right)^{i'i} \,
    +
    \Delta C_\ell^{ij} \,
    \left(\Gamma^{{\rm N}, \, -1}_\ell\right)^{j i'}
    \Delta C_\ell^{i'j'} \,
    \left(\Gamma^{{\rm N}, \, -1}_\ell\right)^{j'i}
    \right]  \\
    \nonumber
    & = & \,
    f_{\rm sky}\, \sum_\ell \, b_{\ell}\, (2\ell+1)\,
    {\rm Tr}
    \left[
    \Delta \mathbf{C}_\ell^{\rm x} \cdot
    \mathbf{\Gamma^{{\rm N},-1}_\ell} \cdot
    \Delta \mathbf{C}_\ell^{\rm x} \cdot
    \mathbf{\Gamma^{{\rm N},-1}_\ell}
    \right]\,,
\end{eqnarray}
where the sum over $\ell$ now denotes a sum over multipole bandpowers of width $b_{\ell}$.
Here $\Delta \mathbf{C}^{\rm x}_\ell \equiv \mathbf{C}_\ell^{\rm x,\rm obs}-\mathbf{C}^{\rm x, \rm th}_\ell(\theta) $,  with $\mathbf{C}_\ell^{\rm x, \rm obs}$ and $\mathbf{C_\ell^{x, \rm th}}(\theta)$ the observed and theoretical cross-angular power spectra (see Fig.~\ref{fig:et2ce_corrl}).
Notice that both $\mathbf{\Delta C^x_\ell}$ and $\mathbf{\Gamma}^{{\rm N} , -1}$ are symmetric, which results in the simplified expression of the second line of Eq.~\eqref{eq:gal_bbh_chi2_matrix}.

\subsection{A fast estimator for
the theoretical redshift-distance cross-correlations} \label{subsec:estimator}
Now, we turn our attention to the challenge of obtaining a theoretical prediction for the
galaxy-BBH cross-correlations in harmonic space, $C_\ell^{ij}$.
The initial approach one might consider is to generate a large number of simulations for each point in the cosmological parameter space ${\theta}$, and then compute the average angular cross spectrum.
While this method is theoretically feasible, it is impractical for statistical inference. Producing and computing the complete set of $\mathbf{C}_\ell$ for a thousand GLASS simulations at a single point in parameter space requires several hours on a modern CPU.

If we wish to probe parameters such as $(H_0, \Omega_m)$ using these observables, we would need to sample approximately $10^4$ points in parameter space, which would require substantial computational resources. Consequently, to compute the log-likelihood in Eq.~\eqref{eq:gal_bbh_chi2_matrix} within a MCMC framework, we need a more efficient method to estimate the theoretical cross-correlations $\mathbf{C}_\ell(\theta)$. We will now demonstrate that this is indeed feasible by exploiting the transformation of the cross-correlations between the tracers $g$ (galaxies) and $b$ (BBHs) as we modify the relationship between redshift ($z$) and luminosity distance ($D_L$).

Notice that the harmonic modes $a_{\ell m}^{b}(D_j)$ of the BBH count map $\Delta N^{b}(D_{j},\vec{\Omega})$ are actually computed by integrating all objects inside that luminosity distance bin, which can be regarded as a sum over smaller distance bins $D_\lambda$:
\begin{equation}
    \label{eq:alm_D}
    a_{\ell m}^{b}(D_i)\,=
    \int_{D_i}  dD
    \,a_{\ell m}^b(D) \quad \rightarrow \quad
    a_{\ell m}^{b}(D_i,\theta)\,=\sum_{D_\lambda \in D_i }\,a_{\ell m}^b(D_\lambda,\theta)\, .
\end{equation}
Suppose that the luminosity distance bins
of our data set are \textit{fixed} -- that is, independent of the cosmology. In practice, this is how we would proceed to compute the cross correlations of a set of observed BBHs with a set of galaxies in fixed redshift bins. Now let's say that the small bins $D_\lambda$ are {\em not} fixed, but rather we fix some set of \textit{very thin redshift bins} $z_\lambda$.  As the connection between redshift and distance is mediated by the cosmological model, we can write $D_\lambda=D_L(z_\lambda,\theta)\,$. Therefore, we can write the harmonic mode in the distance bin $D_i$ as a sum over the redshift bins $z_\lambda$ that fit into that distance bin, for the given cosmology:
\begin{equation}
    \label{eq:alm_D2}
    a_{\ell m}^{b}(D_i,\theta)\,=
    \int_{z \in D_i}  dz
    \,a_{\ell m}^b[D_L(z,\theta)] \quad \rightarrow \quad
    a_{\ell m}^{b}(D_i,\theta)\,=\sum_{z_\lambda \in \, z(D_i,\theta)  }\,a_{\ell m}^b(z_\lambda,\theta)\,.
\end{equation}
The procedure above implements the calculation of the angular power spectra in practice, for the given set of parameters $\theta$.
Now say that we change the set of cosmological parameters from some fiducial ${\theta}_{\rm fid}$ to another arbitrary set ${\theta}$.
Although the amplitudes $a_{\ell m}^{b}(z_\lambda,\theta)$ change, the correlation function still peaks when the redshift of the galaxies coincide with $z_\lambda$. Hence, the main effect of changing the cosmological parameters is the shift in the location of the peak in $D_L(z)$, which can be represented by a remapping of the redshift bins ($z_\lambda$) into the distance bins $D_i$ in Eq.~\eqref{eq:alm_D2}.

Let us then approximate the set of harmonic coefficients in an arbitrary cosmology to those coefficients computed from the distribution of BBHs (and galaxies) in the fiducial model {\em rescaled} to the distances in the new cosmology,
\begin{equation}
    \label{eq:alm_D3}
    a_{\ell m}^{b}(D_i,\theta)\, \equiv \sum_{z_\lambda \in \, z(D_i,\theta)  }\,a_{\ell\,m}^b(z_\lambda,\theta_{\rm fid})\,.
\end{equation}
Now, given that we can compute the angular cross spectrum in very thin redshift bins $z_\lambda$, it is straightforward to show from the equation above that the theoretical angular cross spectrum in arbitrary redshift bins $z_j$ for the galaxies, and distance bins $D_i$ for the BBHs, is given by:
\begin{equation}
\label{eq:alm_D4}
    C^{gb}_\ell(z_j,D_i,\theta) \; =
    \sum_{z_\lambda \in z(D_i,\theta)   }
    \,C^{gb}_\ell(z_j,z_\lambda,\theta_{\rm fid})\,\,,
\end{equation}
where $C^{gb}_\ell(z_j,z_\lambda,\theta_{\rm fid})$ is the cross-angular power spectrum computed in thin redshift bins $z_\lambda$ (both for galaxies and BBHs) and in a fiducial cosmology $\theta_{\rm fid}$, averaged over a given number of realizations (for our purposes, of the order of thousands, as we will see). The only general condition is that the distance bins are wide enough to comprise several thin redshift bins, so that our analysis is not significantly affected by that spatial resolution.

The main reason why Eq. \eqref{eq:alm_D4} is a good approximation derives from the coincidence of the peak of the cross-correlation with the Hubble diagram, which is a generic feature of the large-scale structures and as such should not depend significantly on the galaxy or BBH populations, bias modeling or non-linearities.
The analytical argument is as follows:
for a given set of cosmological parameters $\mathbf{{\theta}}$, the angular power spectrum for the cross-correlations between galaxies and BBHs is given by:
\begin{equation}    C_\ell(z_g,D_L,\mathbf{{\theta}})
=\frac{2}{\pi}
\, b_g(z_g) \, b_{b} (D_L) \,
\int\,dk\,k^2\,
j_\ell\left[k\,\chi(z_g, \mathbf{{\theta}})\right]\,
j_\ell\left[k\,\chi(D_L, \mathbf{{\theta}})\right]\,
P(k,z_g,D_L,\mathbf{{\theta}})\,,
\end{equation}
where $b_g$ and $b_{b}$ are the biases of the galaxies and BBHs, and $\chi$ is the comoving distance.
In the Limber approximation, this turns into
\begin{equation}  C_\ell(z_g,D_L,\mathbf{{\theta}})\approx \frac{\delta\left[\chi(z_g, \mathbf{{\theta}})-\chi(D_L, \mathbf{{\theta}})\right]}{\chi(z_g, \mathbf{{\theta}})^2}
\, b_g(z_g) \, b_{b} (D_L) \,
P\left[k=\frac{\ell+1/2}{\chi(z_g, \mathbf{{\theta}})},z_g,\mathbf{\theta}\right]\,.
\end{equation}
In the equation above, only the cosmological parameters that affect  distance (mainly, $H_0$ and $\Omega_m$) come into play inside the Dirac delta function, while the other parameters (as well as the biases) affect only the amplitude of the spectrum.
Hence, the signal is dictated primarily by the sharp signature of the Dirac delta function, which enforces the match between $\chi(z_g,\mathbf{\theta})$ and $\chi(D_L,\mathbf{\theta})$.

Equation \eqref{eq:alm_D4} is the key approximation that makes our analysis possible.
As we have shown above, the log-likelihood \eqref{eq:gal_bbh_chi2_matrix} will be sensitive mostly to the peak of the cross-correlation signal, broadened by the uncertainty in the BBH localization, and this is overwhelmingly dictated by the distance-redshift relation. The validity of this approximation will be demonstrated in practice, using simulations, in Section~\ref{sec:checks}.

%%%%%%%%%%%%%%%%%%%%%%%%%%%%%
\section{Simulations}
\label{sec:simulations}
%%%%%%%%%%%%%%%%%%%%%%%%%%%%%

In order to assess the potential of the Peak Sirens method to constrain cosmology we forward-model the cosmological distribution of galaxies and BBHs. We will make use of two sets of simulations: the first will be used to compute the theoretical $C_\ell^{gb}$; the second will be used to forecast future data. This first set of simulations is a necessary part of the Peak Sirens method, and must always be performed. The second set is only used to make forecasts, and will be replaced by the real data when it becomes available. Clearly one does not know a priori which is the right cosmology for the first set, but we will show below that the method is largely insensitive to the choice of cosmology.

The starting point is a full-sky cosmological mock of the past light cone, possible thanks to the publicly available \texttt{GLASS} code~\cite{tessore2023glass}. We used \texttt{GLASS} to simulate 1000 lognormal galaxy mocks in very thin redshift bins, of width $dz = 0.02$, over the interval $z \in [0,2]$ -- resulting in the mean redshift bins $\bar z$:
\begin{equation}
\label{eq:zbins}
    \bar z=[0.01,0.03,0.05,...,1.99]\,.
\end{equation}
Each tomographic bin around $\bar z=z_\lambda$ yields a galaxy count map $N^{g}(z_\lambda,\vec\Omega)$, which we then project onto \texttt{HEALPix} maps with resolution \texttt{nside = 256} (that is, $12\times \texttt{nside}^2 = 786432$ pixels). The simulations were created using a fiducial cosmology defined by the following parameters:
\begin{equation} \label{eq:fiducial_cosmology}
H_0=70.0 \,\text{km\,s}^{-1}\,\text{Mpc}^{-1} \; , \quad \Omega_m=0.3 \; , \quad w_0 = -1 \; , \quad w_a = 0 \; .
\end{equation}

From each galaxy catalog and the expected numbers and errors for the GW events, we generate a corresponding BBH catalog, and then compute their correlations. The measures taken to make our simulations more realistic are detailed in the following subsections.

\subsection{Gravitational wave Fisher forecasts}
\label{sec:BHdistribution}

For the construction of realistic BBH catalogs, we first need to assess the expected distribution of source counts and errors in their 3D positions once we know the specifics of the GW detectors, under certain assumptions regarding the true existing population of BBHs, some of which we might observe. Here we considered three networks of GW detectors:
\begin{itemize}
    \item {\bf LVK O5.} The current GW detectors LIGO (both Hanford and Livingston), Virgo and KAGRA, all with design sensitivity \cite{KAGRA:2013rdx}. We considered the 3 years of observations expected for the fifth observing run (O5) currently planned to start in 2027 \cite{obsLVK}, with 80\% duty cycle. This is our near-future scenario only using existing facilities.

    \item {\bf LVK+ET.} The above combined with the triangular-shaped Einstein Telescope (ET), with 10 km arms, located in Italy and with the ET-D sensitivity~\cite{Hild:2010id}.  We assume a total observation time of 5 years, although third generation observatories should be active for a decade, with 80\% duty cycle (i.e. 4 effective years). This is our long-term scenario if current detectors would still be active then, which would help with GW localization.

    \item {\bf ET+2CE.}  The triangular-shaped ET as above, combined with two L-shaped Cosmic Explorers (CE) with 40 km and 20 km arms in the location of the LIGO-Virgo-KAGRA detectors and the baseline sensitivities.\footnote{The CE sensitivities can be found in \url{https://dcc.cosmicexplorer.org/CE-T2000017/public}} We assume again a total of 5 years of observations, with 80\% duty cycle. This is our futuristic best case scenario.
\end{itemize}

The number of BBH mergers per redshift per year with respect to the observer is given by:
\begin{equation}
	\frac{\dd N^b}{\dd z \dd t} = \frac{r(z)}{1+z}\frac{\dd V_c}{\dd z}\,,
\end{equation}
where $V_c$ is the comoving volume and $r(z)$ is the merger rate per comoving volume, as measured by clocks in the source frame -- which is the reason for the additional factor $(1+z)$.
For the merger rate, we assumed the star formation rate \cite{Madau:2014bja} convoluted with a time delay $t_d$, sampled with a lognormal distribution with minimum $t_d = 20 \; \text{Myr}$. Current GW observations are indeed compatible with a merger rate coinciding with the star formation rate at low redshifts, but as we increase our sensitivity to observe high redshift sources in the future, some deviation from this function is expected to be observed, due to the time it takes to a binary to coalesce once it is formed. The normalization of the merger rate is the parameter that has greatest impact on the total amount of events prospected. We used the upper $1\sigma$ limit from the measurement of LVK GWTC-3 at $z = 0.2$, where the highest precision in $r$ is achieved:
\begin{equation}
	r(0.2) = 28^{+14}_{-9} \; \text{Gpc}^{-3} \text{yr}^{-1}\,.
\end{equation}
This gives around 41200 BBHs per year up to $z \sim 10$ and 24000 up to $z = 2$.

For the mass distribution, we assumed the \textsc{power law + peak} model \cite{LIGOScientific:2020kqk}, whose parameter values were set to the best fits from LVK GWTC-3 observations \cite{LIGOScientific:2021psn}. We considered an additional extension of this mass model to include the possible intermediate mass black-holes that cannot be seen with the current GW detectors but are in the horizon for the next generation. This extension is simply a lognormal distribution from 88 $M_{\odot}$ to 1000 $M_{\odot}$. For the spins of each black hole, we assumed their components along the binary's angular momentum, $s_{1z}, s_{2z}$, to follow the \textsc{default} distribution, again with parameter values from~\cite{LIGOScientific:2021psn}, while the transverse components were set to vanish, avoiding the precession terms in the waveform for simplicity. Finally, the BBH positions $(\theta, \phi)$ injected were uniformly sampled in the sky area, while the polarization angle $\psi$ and the inclination angle of the binary's angular momentum $\iota$ with respect to the line of sight were sampled from a uniform distribution in their cosines.

We performed Fisher forecasts using the code \texttt{GWDALI} \cite{deSouza:2023ozp},\footnote{Here we employ only the Fisher matrix and not the full DALI method with doublet or triplet~\cite{Sellentin:2014zta}.} with the following 11 free parameters:
\begin{equation}
    (M_c, \eta, D_L, \iota, \theta, \phi, \psi, s_{1z}, s_{2z}, t_c, \phi_c)\,,
\end{equation}
where $M_c$ is the redshifted chirp mass, $\eta$ is the symmetric mass ratio, and $t_c$ and $\phi_c$ are the time and phase of coalescence. We considered as threshold for detection having the total SNR of the network greater than 12. For the LVK+ET network, we have further imposed that at least one of the LVK detectors have SNR > 1. That does not mean, therefore, that the events included in this case would actually be detected by the 2nd generation network. Rather, LVK would help a 3rd generation detector (here, the triangular ET) to triangulate the signal improving localization knowledge, similarly to how Virgo has helped LIGO on localizing sources in their first three observing runs  even if for the events of which Virgo alone can not claim a detection.  After these cuts in SNR, we prospect the total amount of 377 observed events for LVK O5, while for LVK+ET and ET+2CE we expect this number to grow to, respectively, 54991 and 73260 detected events.

Fisher matrix based inference for GW parameter estimation works well for most high SNR events. However, the Gaussian approximation of the exact GW likelihood breaks down in some specific cases. Notably, for low inclinations of the binary's angular momentum, $\iota$ and distance become degenerate and the Fisher is difficult to invert \citep{Chassande-Mottin:2019nnz,Mancarella:2024qle}. To avoid this issue, like other codes, \texttt{GWDALI} uses the singular-value decomposition and invert the Fisher matrix removing its small eigenvalues, defined by a threshold $r_{\rm cond}$. We chose this parameter to be the smallest value possible, $r_{\rm cond} = 10^{-16}$. Furthermore, in order to be conservative and ensure our results are accurate, we additionally excluded from our analysis the events with $\iota < 5\degree$ or $\iota > 175\degree$, where the degeneracy is crucial. Moreover, another caveat of the Fisher approximation is related to the sky position uncertainty with single detectors. A network of detectors allows improved sky localization due to the time-delay between spatially separated detectors. For a single detector or co-located detectors, however, the most likely sky location is a multiply connected region in the sky, for which a Gaussian approximation breaks down. This is why we chose not to forecast ET or CE alone.

\begin{figure}[t]
    \centering
    \includegraphics[width=\linewidth]{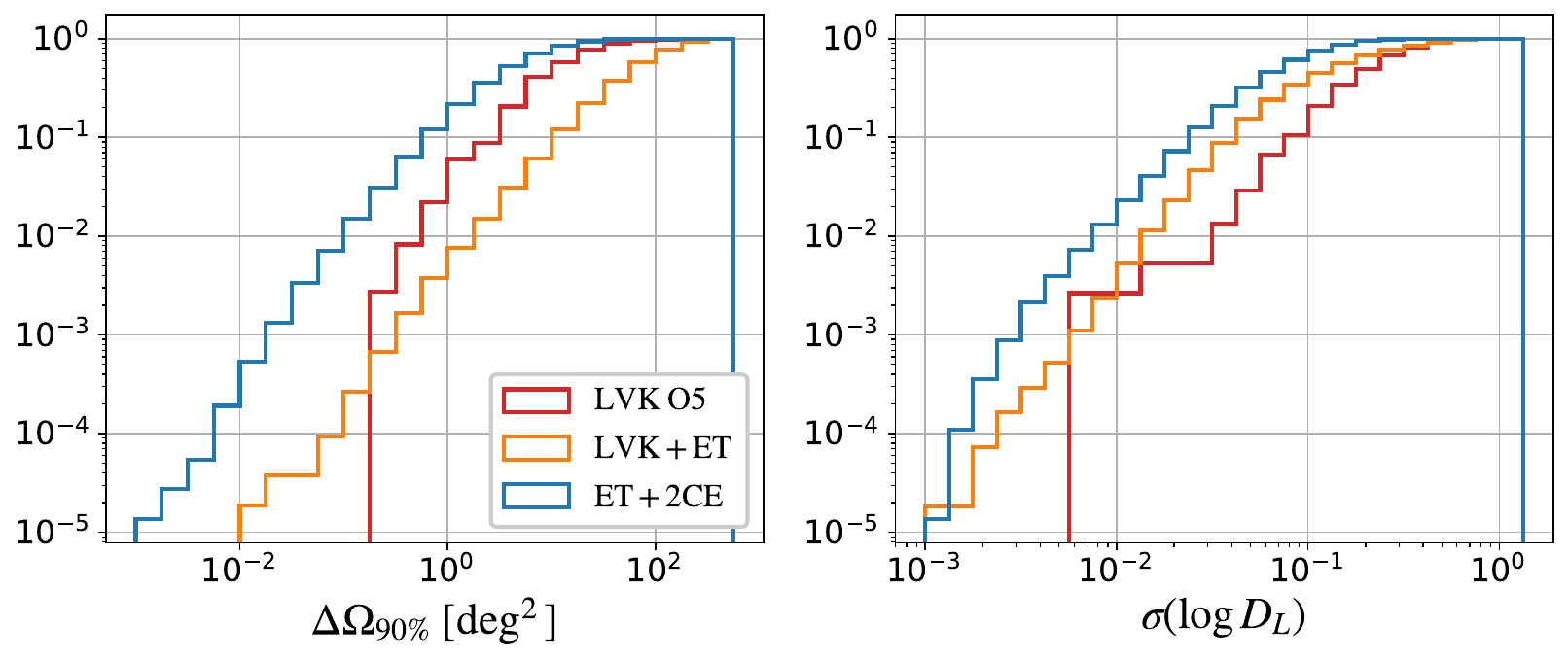}    \caption{Cumulative distribution of forecast sky position uncertainties (left) and relative distance errors (right) for the three GW detector networks here considered. The number of events is much higher for LVK+ET compared to LVK O5, especially at high $z$, a fraction of which with poor localization, result in a modest improvement in the average localization.
    See also Figure~\ref{fig:bbh_catalogs2}.}
    \label{fig:bbh_catalogs}
\end{figure}

The waveform model considered was the \texttt{IMRPhenomHM} \cite{London:2017bcn, Kalaghatgi:2019log}, implemented with the help of the \texttt{LAL} package \cite{lalsuite}, which includes higher modes that help breaking the degeneracy between $\iota$ and distance. We included in the current public version of \texttt{GWDALI} the effect of the Earth's rotation, both in the antenna pattern functions that modulate the amplitude of the waveform, and in the time delay affecting the phase. This effect mildly helps decreasing the sky area uncertainty, and can be more relevant for low mass events. Fig.~\ref{fig:bbh_catalogs} shows the resulting distribution of the errors on the distances and 90\% C.L. sky area, which can be written as
\begin{equation}
    \Delta\Omega = 2\pi \ln(10)|\sin\theta|\sqrt{(F^{-1})_{\theta \theta}(F^{-1})_{\phi \phi} - [(F^{-1})_{\theta \phi}]^2}\,,
\end{equation}
where $F$ is the GW Fisher matrix for the network and our chosen waveform model. We find reasonable agreement with the forecasts of \cite{Iacovelli:2022bbs} for BBHs.

\begin{figure}[t]
    \centering
\includegraphics[width=\linewidth]{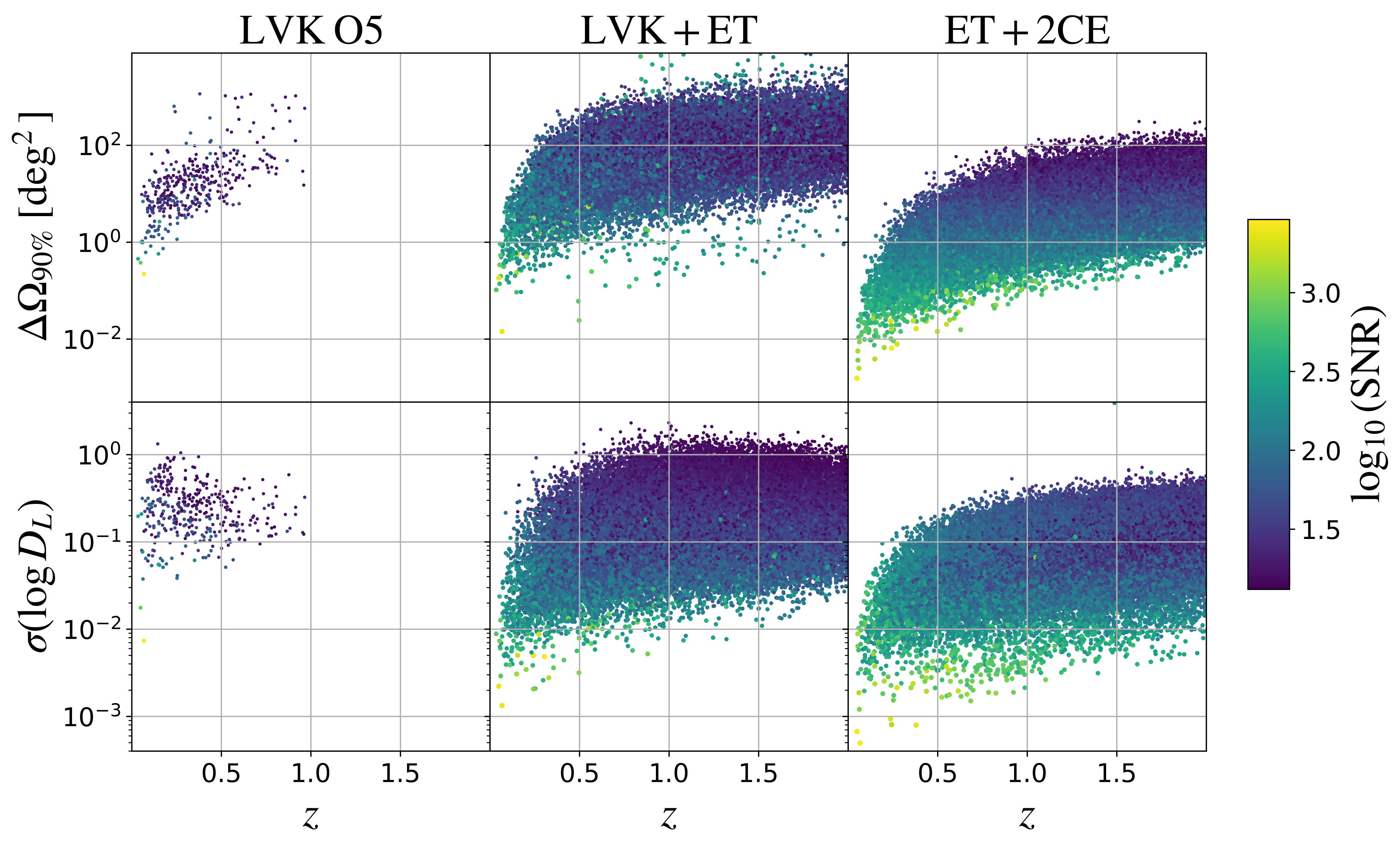}
    \caption{Distribution of the errors in the angular position of the events (upper panels) and distance errors (lower panels), as functions of redshift, for the three networks of detectors considered. The color bar shows the total SNR of the network -- which, as expected, anti-correlates with the errors.
    We also note that, for the ET+LVK case, the sky area depends mostly on the SNR of the 2g network, indicating that the triangulation with the 2g detectors would be crucial to improve sky localization compared with a single 3g detector.
    }
    \label{fig:bbh_catalogs2}
\end{figure}

Finally, we have also studied the network composed by two L-shaped Einstein Telescopes with 15 km arms, which is an alternative configuration being considered for ET, combined with the two Cosmic Explorers. We found no substantial difference in the sky areas and distance errors as compared to ET+2CE.
The resulting distribution of BBH errors in the three scenarios are shown in Fig.~\ref{fig:bbh_catalogs2}, where we limited the number of events for the LVK+ET and ET+2CE networks to 3000 for visualization purposes.
Figs. \ref{fig:bbh_catalogs} and \ref{fig:bbh_catalogs2} also serve the purpose of showing that redshift-space distortions  are washed out by the large distance errors in the BBH distances: peculiar velocities are typically smaller than $\sigma_v \lesssim 10^3 \, {\rm km} \, {\rm s}^{-1}$, corresponding to $\sigma_z \lesssim 0.3 \%$, while the typical BBH distance errors are $\gtrsim 1 \%$ even for ET+2CE.
Therefore, at the scales we are interested, redshift-space distortions are at most a correction to the bias of the galaxies and the BBH hosts.

\subsection{Galaxy and BBH mocks}
\label{sec:gal_bbh_mocks}
Given the BBH histograms for the area and distance errors from Section \ref{sec:BHdistribution}, we have now two tasks in order to put the cross-correlation analysis to test: the first is to create galaxy mock catalogs simulating stage-IV surveys. The second task is to populate these galaxy mocks with BBH events, informed by the histograms shown in Fig.~\ref{fig:bbh_catalogs}.

Since the number of observed galaxies far exceeds even the most optimistic projections for GW detections, the limiting factor for the signal is expected to be primarily the number of BBHs and the uncertainties in their 3D positions. This was confirmed by analyzing the signal from maps with a substantially larger number of galaxies, where the results remained effectively unchanged. Thus, the precise observed galaxy distribution is not critical for this analysis.
Consequently, rather than following a specific survey, we took all redshifts as spectroscopic and adopted the Smail formula for the redshift distribution of the observed number of galaxies~\cite{Smail_1994}:
\begin{equation}
    p(z)\sim\left(\frac{z}{z_0}\right)^\alpha\,\exp\left[-\left(\frac{z}{z_0}\right)^{\beta}\right]\,,
\end{equation}
with $z_0=0.7$, $\alpha=2.0$, $\beta=1.5$, and a linear galaxy bias, $b_g(z) = (1+z)$.
This prescription yields a good fit for magnitude-limited optical surveys, as it captures in a simple formula volume effects at low redshifts, the evolution of the galaxy luminosity function, as well as decreasing counts due to increasingly fainter objects at high redshifts.
With our choice of parameters (including the normalization of the Smail distribution) we obtained approximately 34 million galaxies up to $z=2$, a number that will easily be surpassed by upcoming surveys. To account for partial sky coverage, we apply a mask to the galaxy maps, restricting the analysis to latitudes \(\vert \phi \vert \geq 19^\circ\), approximately \sfrac{1}{3} of the sky.

Now, we are left with the task of bringing the detection prospects (given by the distance and angular position, alongside its error histograms supplied by \texttt{GWDALI}) to the same universe as our galaxy mocks. From the galaxy mock catalogs, we (i) select a fraction of them to host a GW event (and strike that galaxy out of the galaxy sample), and (ii) attribute a luminosity distance and angular error to each event following the steps discussed in the next section.

\subsubsection{BBH angular distribution}
\hypertarget{step1}{}
First, we want to select galaxies as hosts of BBHs by accounting for both the galaxy distribution itself and for the selection function of the GW detector network in the sky (even taking into account Earth's rotation, existing or planned networks will not cover the whole sky evenly).
To accomplish this, we created 2D histograms for each redshift bin, representing the events that meet the detection threshold outlined in the forecasts discussed in the previous section. These histograms will assist us in selecting GW hosts for only the {\it observed} BBHs.

The angular positions of BBHs in our forecasts are represented in GPS coordinates (longitudes and latitudes with respect to the Greenwich meridian and the Earth's equator), since all GW detectors are terrestrial. To better construct the histograms, we create multiple copies of each event, to which we assign random right ascension coordinates, accounting for the random detection times. Then, we convert these to galactic coordinates \texttt{(LAT,LON)}. Converting its distances to redshift through the fiducial cosmology \eqref{eq:fiducial_cosmology}, we project these events into four redshift bins $\bar{\bar z}$ of width $\Delta z=0.25$:
\begin{equation}
    \label{eq:thick_zbins}
    \bar{\bar z}=[0.25,0.75,1.25,1.75]\,\,.
\end{equation}

For each of these, we construct event histograms as \texttt{HEALPix} maps of \texttt{nside=32}. The lower resolution in redshift and position is justified by the fact that these distributions vary slowly in redshift, together with their smooth distributions in the sky. Smoothing the generated BBH histogram for each of the four $j$ bins results in the angular distribution of observed BBHs $h_{j}(p)$
(where $p$ denotes a pixel in a \texttt{HEALPix} map of \texttt{nside=256}). Multiplying our galaxy count maps $N^g_i(p)$ in each of the thin redshift bins defined in Eq.~\eqref{eq:zbins} with the GW selection function $h_{j}(p)$ in the corresponding thick bin $j$ yields a `joint BBH PDF' from which we will draw our BBH positions.

\begin{figure}
  \makebox[\textwidth][c]{\includegraphics[width=1\linewidth,clip]{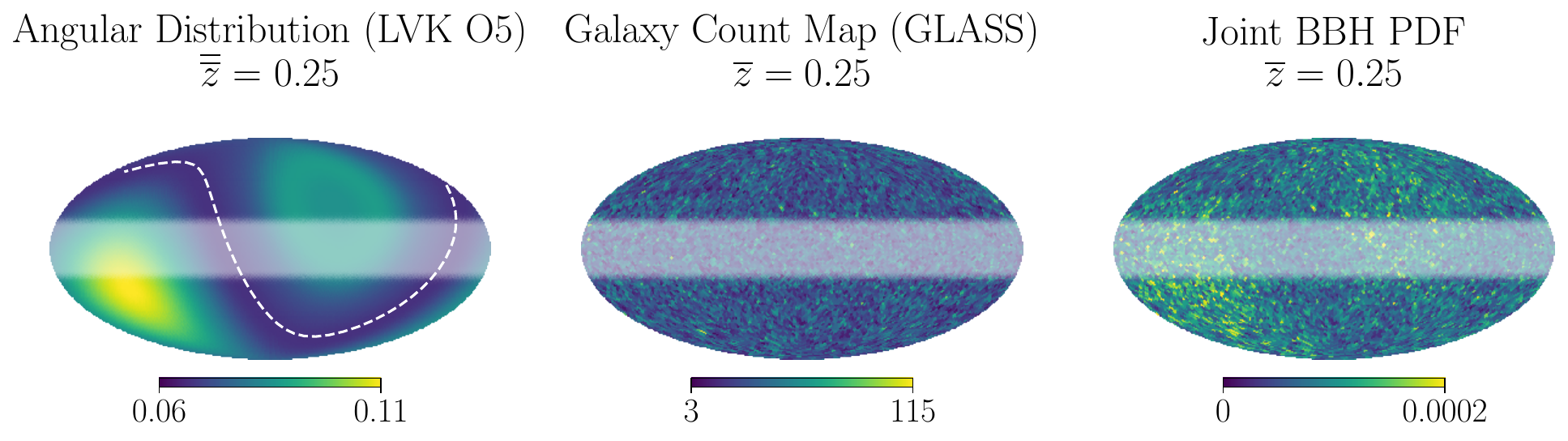}}%
  \caption{\emph{Left:} Angular distribution obtained from a 2D histogram of GW events in the sky for LVK O5. The white, dashed line represents the celestial equator plane, while the white-shaded region represents the Milky Way,  which we will mask from our galaxy catalog. \emph{Center:} Galaxy count map given by a \texttt{GLASS} mock catalog at redshift bin $\bar z=0.25$. \emph{Right:} Joint BBH PDFs taking into account the galaxy mock distributions and the angular distributions of BBH sensitivities of each detector  network.}
  \label{fig:joint_bbh_pdf}
\end{figure}

Fig.~\ref{fig:joint_bbh_pdf} illustrates the steps necessary to construct this PDF. The left panel displays the smoothed selection function for the LVK O5 network, for the first redshift bin (centered in $\bar{\bar{z}}=0.25$), where 77\% of all observed events lie. It is interesting to note that this network is indeed sensible to selection effects, showing a noticeable lack of events along the celestial equator. This pattern arises due to the interferometric GW detectors' greater sensitivity to waves that arrive perpendicular to the detector plane, which is evident in the figure as all three detectors (LIGO, Virgo, and KAGRA) are positioned in the Northern Hemisphere. In contrast, for the third-generation network, the distribution of observed sources is expected to be more uniform across the sky, as the GW signals from most stellar-mass binaries should surpass the detection threshold, regardless of their location. To select the exact positions of each BBH event, we first need an underlying matter field -- or, more simply, a galaxy count map. This is where we make the connection between the galaxy mock from \texttt{GLASS} (center of Fig.~\ref{fig:joint_bbh_pdf}) and the redshift and angular distributions from \texttt{GWDALI} -- creating the joint PDF (right image of the figure), and enabling us to create a corresponding mock BBH catalog.

\subsubsection{BBH position errors}
\hypertarget{step2}{}
After drawing the true host galaxies for our GW sources, we know their \texttt{HEALPix} pixel (or \texttt{LAT,LON} in galactic coordinates) and redshift bin, but we still need to associate an observed distance and angular error to each of them. What we would like to do is to map the list of observed BBHs in our forecasts of section \ref{sec:BHdistribution}, for which we know the errors in 3D position (but the injected angles are just uniform in the sky), onto the physically relevant BBH map sampled from the product of the galaxy distribution with the GW selection function.

To properly attribute a position error in galactic coordinates that depends on redshift and sky position (but smoothly, across large scales), we again distribute the BBHs of the GW forecasts on a second set of histograms (or maps) $h_{{\rm geo},j}(p)$, but now in geocentric coordinates, where $j$ are the four redshift bins defined by \eqref{eq:thick_zbins}, and $p$ are \texttt{HEALPix} pixels in a map of \texttt{nside=4}. Then, for each of the $N^b_i$ BBHs drawn in the thin redshift bin $i$ in our physically relevant maps, we convert its coordinates to geocentric, evaluate in which thick redshift $j$ bin and lower \texttt{nside} $p$ pixel it belongs to, and associate to it a random BBH in $h_{{\rm geo},j}(p)$, from which we can
 get the errors in distance and sky position.

In short, say that one event inside $h_{{\rm geo},j}(p)$ has luminosity distance error $\Delta D_L (j, p)$ and geocentric angular errors $\Delta \theta(j,p)$ , $\Delta \phi(j,p)$. Then the attributed luminosity distance of a BBH mock event on thin redshift bin $i$ and \texttt{nside=256} pixel $p$ is

\begin{equation}
\label{eq:bbh_newdist}
    D_i + \mathcal{N} \left[ 0,\Delta D_L (j,p) \right]
\end{equation}
with
\begin{equation}
    D_i\equiv\mathcal{U} \left[ D_L(\bar z_i-dz/2) \, , \, D_L(\bar z_i+dz/2) \right]\,\,,
\end{equation}
where $\bar z_i$ is the mean redshift of the bin $i$, $\mathcal{U}(a,b)$ is an uniform probability distribution from a to b and $\mathcal{N}(\mu,\sigma)$ is the normal distribution. This distance is finally converted back to redshift, and the event is associated to another thin redshift bin according to this new redshift.

The new (geocentric) angular position \texttt{(RA,DEC)} is given by, respectively,
\begin{align}
    &\mathcal{N}\left[ \theta_{{\rm geo},k},\Delta \theta (j,p) \right]\,\,,\\
    &\mathcal{N}\left[ \phi_{{\rm geo},k},\Delta \phi (j,p) \right]\,\,,
    \label{eq:bbh_newpos}
\end{align}
where $\theta_{{\rm geo},k}$ and $\phi_{{\rm geo},k}$ are the geocentric coordinates (converted from the $p$-th pixel in \texttt{HEALPix} galactic coordinates). Finally, these coordinates are brought back to galactic, and the event is associated to another pixel of \texttt{nside=256} according to its new position. This way we can realistically account for the distribution of errors in distance and position in the sky and their correlations.
We found that including position-dependent errors in the 3D localization of the BBHs has an impact in the noise of the cross-correlations, worsening the constraints by a small factor.

\begin{figure}[t]
    \centering
    \includegraphics[width=1\linewidth,trim={3.5cm 1.3cm 3.0cm 0}, clip]{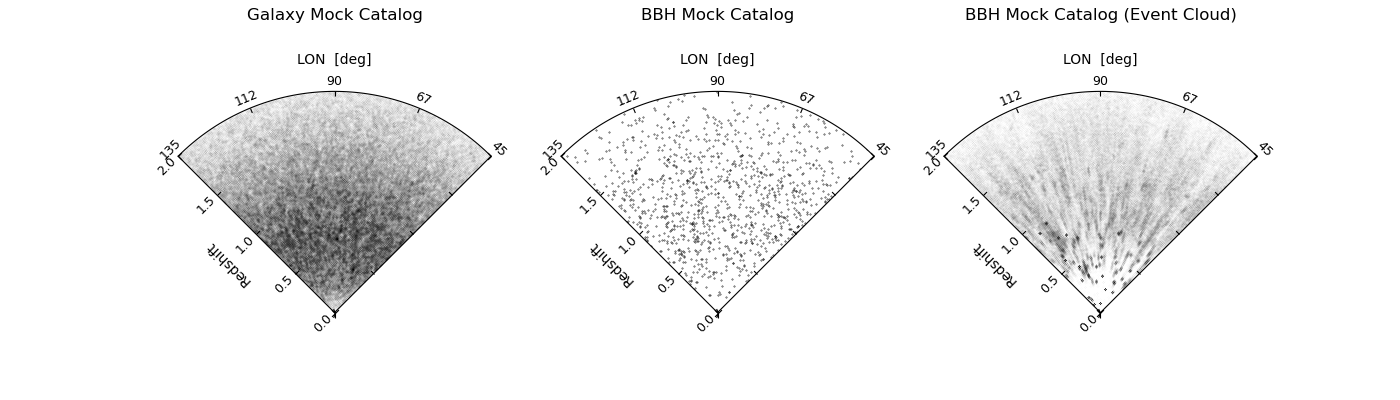}
    \caption{A galaxy-BBH mock catalog pair for the ET+2CE network by \texttt{GLASS}.
    The three panels show the same slice in a cone of latitude cone of 3 degrees, for galaxies (left), BBHs in their true positions (center) and BBHs after including radial distance and angular position errors (right).
    The right plot shows the BBH catalog after applying the \textit{event cloud}, where we re-sampled each BBH event 100 times.
    From this figure it can be seen that low-redshift events are more well-resolved, whereas distant events tend to be more spread out over a ``cloud'' of possible positions. }
    \label{fig:mock_lightcones}
\end{figure}

Given these uncertainties in the radial distances and angular positions attributed the BBHs, there is considerable statistical noise in the cross-spectrum $C_\ell^{gb}(z_g,z_b)$.
To mitigate this issue and achieve a better sampling of the error distributions, we implement an \textit{event cloud} approach, re-sampling $N$ times each event by drawing from the assumed distributions for their angular and radial positions given the corresponding uncertainties, according to Eqs.~\eqref{eq:bbh_newdist}-\eqref{eq:bbh_newpos}. Fig.~\ref{fig:mock_lightcones} illustrates a slice of one of our simulations, with a galaxy  map (left panel), the true position of the BBHs (center) and the event clouds representing the uncertainties in the BBH positions with an event cloud for each object (right panel).

With this approach we are now ready to compute the harmonic cross-correlation function with \eqref{eqn:cros_corr}, and ultimately the cross-angular power spectrum \eqref{eqn:angpow}.
However, there is an addition step we can take, which is to account for the magnification effect on the signal from the BBHs by the intervening large-scale structure.

%%%%%%%%%%%%%%%%%%%%%%%%%%%%%%%%%%%%%%%%%%%%%%%
\subsubsection{Weak lensing effects}
\label{sec:weaklensing}

The importance of lensing for GW correlations was first realized by~\cite{oguri2016measuring} -- see also \cite{2013PhRvL.110o1103C} for a related application. Fortunately, in addition to the density map and tracers on the past light cone, \texttt{GLASS} also provides the convergence map $\kappa_i(p)$ for each source redshift plane $z_i$ and \texttt{HEALPix} pixel $p$:
\begin{equation}
    \kappa_i(p)=\frac{3\Omega_m H_0^2}{2}\,\int_{0}^{z_i}\,\delta \left[ D_c(z'),\hat{\Omega}_p \right] \frac{D_c(z')\,D_c(z',z_i)}{D_c(z_i)}\frac{1+z'}{H(z')}\,\dd z'\,\,,
\end{equation}
where $D_c$ is the comoving distance.  The convergence field changes the magnification of source dark sirens, which translates into a shift in the luminosity distance~\cite{2013PhRvL.110o1103C,Oguri_2018}:
\begin{equation}
\label{eq:bbh_magnification}
    D_i'(p)=\frac{D_i}{\sqrt{\mu_i(p)}}\,.
\end{equation}
We will neglect the shear and use the following  approximation for the magnification $\mu$, which is valid except in the strong-lensing regime~\cite{Takahashi:2011qd}:
\begin{equation}
    \mu_i(p)=\big[1-\kappa_i(p)\big]^{-2} \; .
\end{equation}

Note that \texttt{GLASS} provides the mean convergence over the pixel area, which will be denoted as $\bar{\kappa}_i(p)$ from now on. We are interested in the real convergence $\kappa_i(p)$ suffered by the source dark siren, which can be obtained from a probability distribution function of $\kappa$. Several methods exist to fit the convergence PDF over high-resolution pixels (on the order of a few arcseconds; see e.g.~\cite{Das_2006,Alfradique:2024fkb}). A log-normal fit for the convergence PDF has been established for our redshift intervals of interest $(0 < z < 3)$:
\begin{equation}
\label{eq:kappa_pdf}
    f_\kappa(\kappa)=\frac{\exp\left[-\frac{[\log(\kappa+\mu_{\rm logn}-\bar\kappa)-\mu_{\rm gau}]^2}{2\sigma_{\rm gau}^2}\right]}{\sqrt{2\pi}\sigma_{\rm gau}\,(\kappa+\mu_{\rm logn}-\bar\kappa)}\,,
\end{equation}
where the shape parameters $\mu_{\rm logn}$, $\mu_{\rm gau}$ and $\sigma_{\rm gau}$ are functions of $z$, $\Omega_m$, and $\sigma_8$, and $\bar{\kappa}$ is set to ensure a zero mean to $f_{\kappa}$~\cite{Marra:2013roi}.

The crucial property of these lensing PDFs, for what we are concerned, is their asymmetry, which increases with redshift. In particular, their peak at negative values translates into a shift of the peak of our correlation function, and the peak position in the Hubble diagram is the cornerstone of our proposed methodology.

%%%%%%%%%%%%%%%%%%%%%%%%%%%%%%%%%%%%%%%%%%%%%%%
%\newpage
%\FloatBarrier
%%%%%%%%%%%%%%%%%%%%%%%%%%%%%

\begin{figure}
    \centering
    \includegraphics[width=0.9\linewidth]{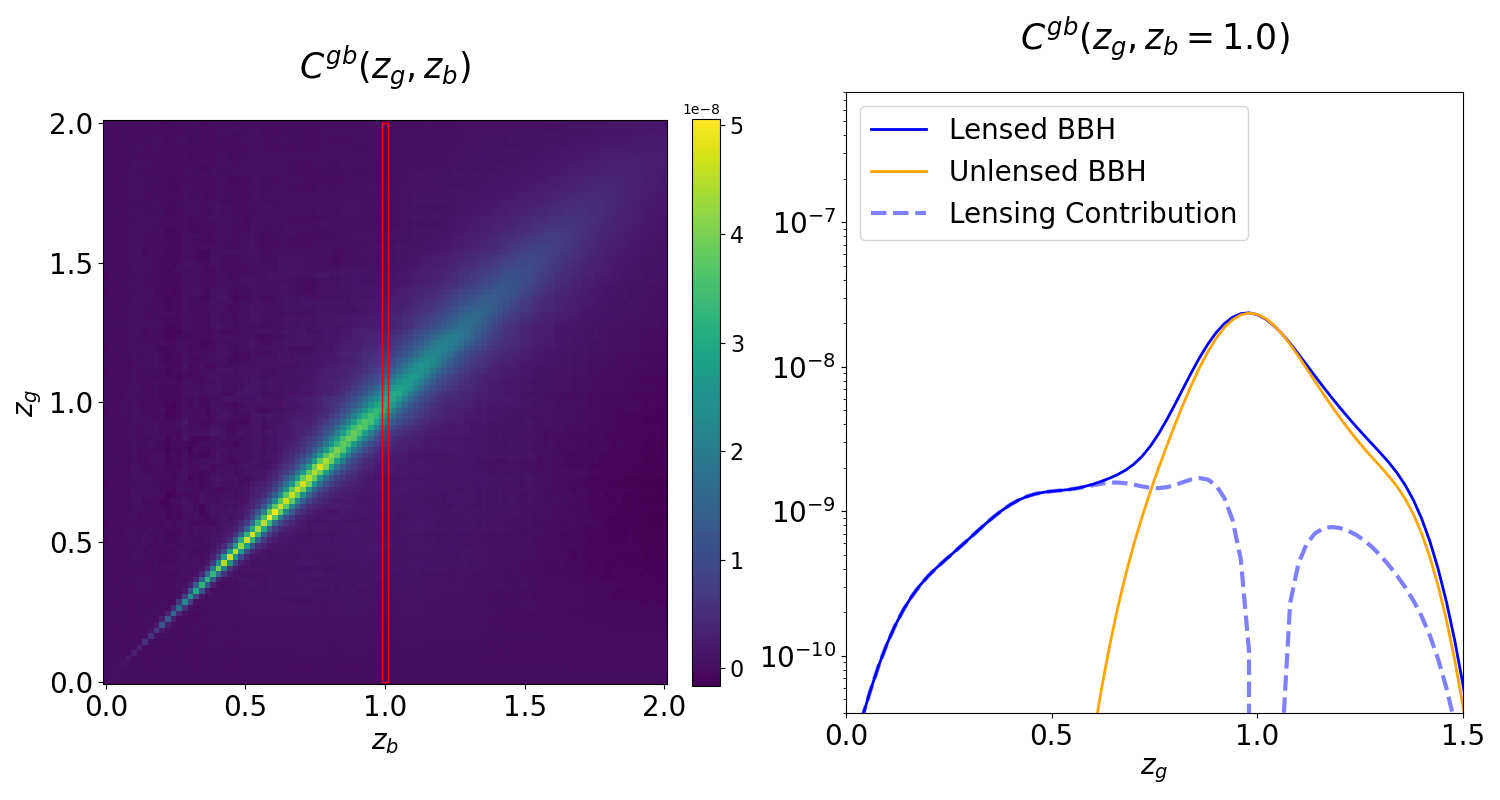}
    \caption{
    \emph{Left:} Line-of-sight cross-correlation function $C^{gb}(z_g,z_b)$ for $\ell_{\text{max}}=100$ -- see Eq.~\protect\eqref{eqn:lineofsight_corr}.
    \emph{Right:} Cross-correlation profile  $C^{gb}(z_g,z_b=1.0)$ (red rectangle in the left panel), with (black solid line) and without (blue dotted line) the lensing magnification contribution (which is shown as the red dashed line).
    Magnification boosts the left side of the peak ($z_g < z_b$), while de-magnification enhances very slightly the correlations on the right side -- see Eq.~\eqref{eq:bbh_magnification}.
    }
    \label{fig:lensed_vs_unlensed}
\end{figure}

We have taken into account GW lensing in our simulations following the steps outlined above. For visualization purposes, we show the redshift-space correlation function \eqref{eqn:RealSpaceCorr} along the line-of-sight (that is, taking pairs separated only along the radial direction, $\hat{r}_i\cdot\hat{r}_j=1$), which is the sum of all angular power spectra from $\ell=2$ up until an arbitrary $\ell=\ell_{\rm max}$:
\begin{equation}
    \label{eqn:lineofsight_corr}
    C^{gb}(z_g,z_b)\equiv\sum_{\ell=2}^{\ell_{\rm max}}\,\frac{(2\ell+1)}{4\pi}\, C_{\ell}^{gb} (z_g,z_b)\,\,.
\end{equation}
Fig.~\ref{fig:lensed_vs_unlensed} shows, on the left, the mean correlation matrix $C^{gb}(z_g,z_b)$.  The right panel shows the profile of that correlation matrix taken along $z_b=1.0$, with and without the effect of lensing of the GWs.
The lensing magnification contribution increases the correlation signal at $z_g < z_b$, shifting slightly the peak of the correlation to lower $z_g$. The Peak Sirens method is not significantly affected by the long tail of $C^{gb}(z_g,z_b)$ for $z_g < z_b$ produced by magnification, but the subtle shift in the peak, if not properly taken into account, can bias the determination of cosmological parameters in future-generation probes -- see Section 4 and the right plot of Fig.~\ref{fig:halofit_lens_biases}.

Fig.~\ref{fig:lensed_vs_unlensed} can also be directly compared with the analytical prediction of Ref.~\cite{oguri2016measuring} -- see Fig.~1 of that paper. Apart from an overall normalization, the results for the weak-lensing contribution to the signal of $C_\ell^{gb}$ for the optimal case of ET+2CE are consistent with that analytical estimate.
However, due to the long tail on the convergence PDFs, we observe a non-negligible correlation between BBHs and galaxies at redshifts far below $z_g=1$.
One of the main differences between our results and those of Ref. \cite{oguri2016measuring} is the steep drop of the lensing signal as  $z\to 0$, which is due to the fact that the convergence PDFs become narrower and nearly Gaussian as we $z \to 0$, corresponding to a fast decreasing fraction of highly magnified objects.

%???
%\FloatBarrier
\section{Results}
\label{sec:results}

\subsection{Forecasts for cosmological parameters}

The data used in this analysis, denoted as $\mathbf{C}_\ell^{\text{obs}}$, consists of the average cross-correlations from 1000 simulations computed in our fiducial cosmology. As discussed in subsection \ref{subsec:estimator}, since actual BBH data is observed in terms of luminosity distance, it is necessary to convert the cross-correlations from $C^{gb}_\ell(z_g, z_b)$ to $C^{gb}_\ell(z, D_L)$. The theoretical input we use in the likelihood is the average of the same simulations, translated to any given set of cosmological parameters via Eq.~\eqref{eq:alm_D4}.

The distance bins are kept fixed but can be defined arbitrarily, independently of the redshift bins. For this analysis we choose 50 distance bins ranging from $D_{\text{fid}}(z=0)$ to $D_{\text{fid}}(z=2)$, in equally spaced redshift intervals, where by $D_{\text{fid}}$ we mean the luminosity distance calculated in the fiducial cosmology. This binning strategy offers  the practical advantage that it ensures that each distance bin includes one or more thin redshift bins in the discrete sum of Eq.~\eqref{eq:alm_D4}.

\begin{figure}[t]
    \centering
    \includegraphics[width=0.55\linewidth]{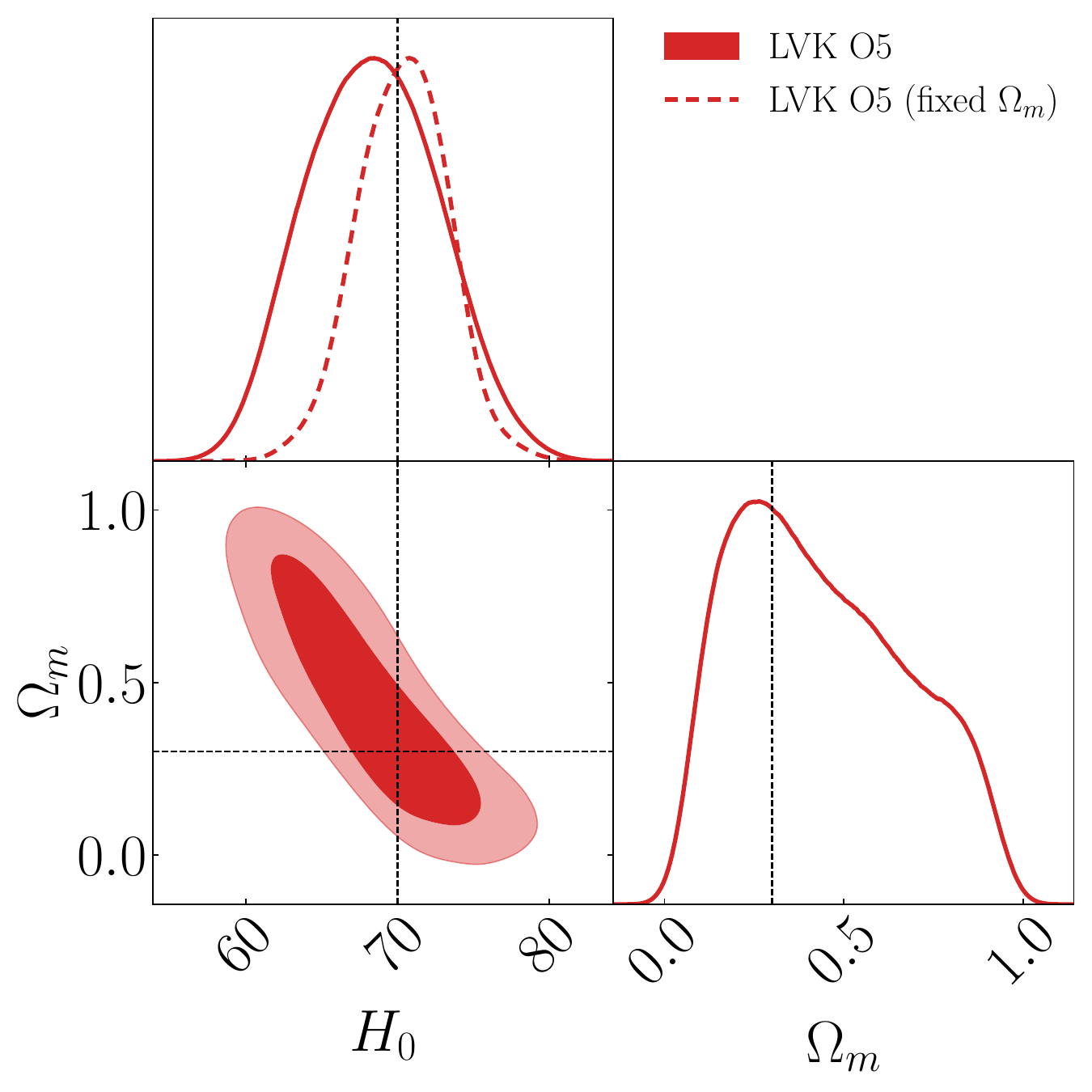}
    \caption{Posteriors for $\Lambda$CDM for the LIGO-Virgo-KAGRA case. The dashed PDF corresponds to the posterior on $H_0$ obtained by fixing $\Omega_m$ to its fiducial value.}
    \label{fig:lvko5}
\end{figure}

For our short-term LVK O5 forecast we account for the effects of weak lensing magnification on luminosity distances, the impact of non-linear corrections to the matter power spectrum using HaloFit \cite{2003MNRAS.341.1311S}, and a visibility mask covering \sfrac{1}{3} of the sky applied to the galaxy maps -- see Fig.~\ref{fig:joint_bbh_pdf}. Since the cross-correlation signal for LVK O5 is significantly weaker compared to ET+2CE, we limit the likelihood analysis to the redshift range $0 \leq z \leq 1$, and to multipoles $2 \leq \ell \leq 400$.
Additionally, we employ broad uninformative priors for \(H_0\) and \(\Omega_m\):
\begin{equation}
    H_0 \sim \mathcal{U}[40, 100]\, \text{km}\, \text{s}^{-1}\, \text{Mpc}^{-1} \, , \quad \Omega_m \sim \mathcal{U}[0.01, 0.99] \, .
\end{equation}

The results from this scenario, depicted in Fig.~\ref{fig:lvko5}, yield \(H_0 = 68.3^{+4.2}_{-5.0} \, \hunit \), and \(\Omega_m = 0.44^{+0.15}_{-0.37}\) when we try to constrain these two parameters jointly. If, instead, we fix $\Omega_m$ and vary only $H_0$, as often done in current dark siren analyses, we obtain $H_0 =70.4 \pm 3.0\,\hunit$, which significantly  outperforms the constraints obtained with the GWTC-3 \cite{GWTC3_constraints,Alfradique:2023giv} dataset which have \(\sigma(H_0) \simeq 15\, \hunit \). The forecast constraint on $\Omega_m$ is however very weak for LVK O5.

For middle- and long-term scenarios, we investigate LVK+ET and ET+2CE as potential networks of GW detectors. For these forecasts we also consider, besides $\Lambda$CDM, a second  model, $w_0w_a$CDM, where the equation of state is allowed to vary.
We adopt narrower, but still uninformative, flat priors on $H_0$ and $\Omega_m$.
For $w_0$ and $w_a$, they are either fixed or allowed to vary within our uniform prior ranges. The priors are:
\begin{eqnarray}
    H_0 \sim \mathcal{U}[65, 75]\,\hunit , \; \Omega_m \sim \mathcal{U}[0.1, 0.5],\;
    w_0 \sim \mathcal{U}[-1.5, -0.5]  , \; w_a \sim \mathcal{U}[-5, 2]
\end{eqnarray}

\begin{figure}%[t!]
    \centering
    \includegraphics[width=0.55\columnwidth]{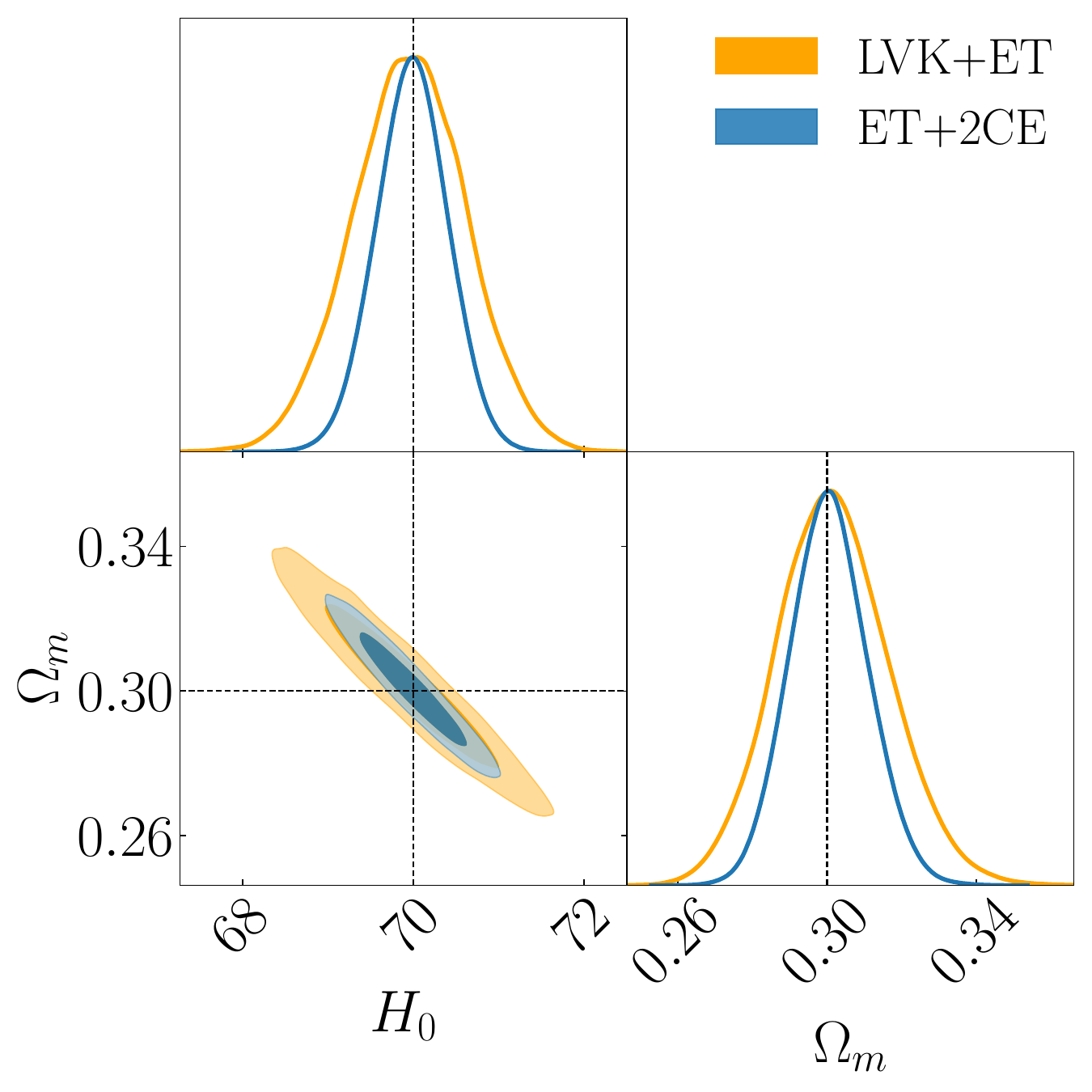}
    \caption{Posteriors  for $\Lambda$CDM for two sets of GW networks: Einstein Telescope + 2 Cosmic Explorers (blue) and LIGO-Virgo-KAGRA + Einstein Telescope (yellow). See also Table~\ref{tab:results}.}
    \label{fig:et2ce}
\end{figure}

\begin{figure}%[t!]
    \centering
    \includegraphics[width=0.67\columnwidth]{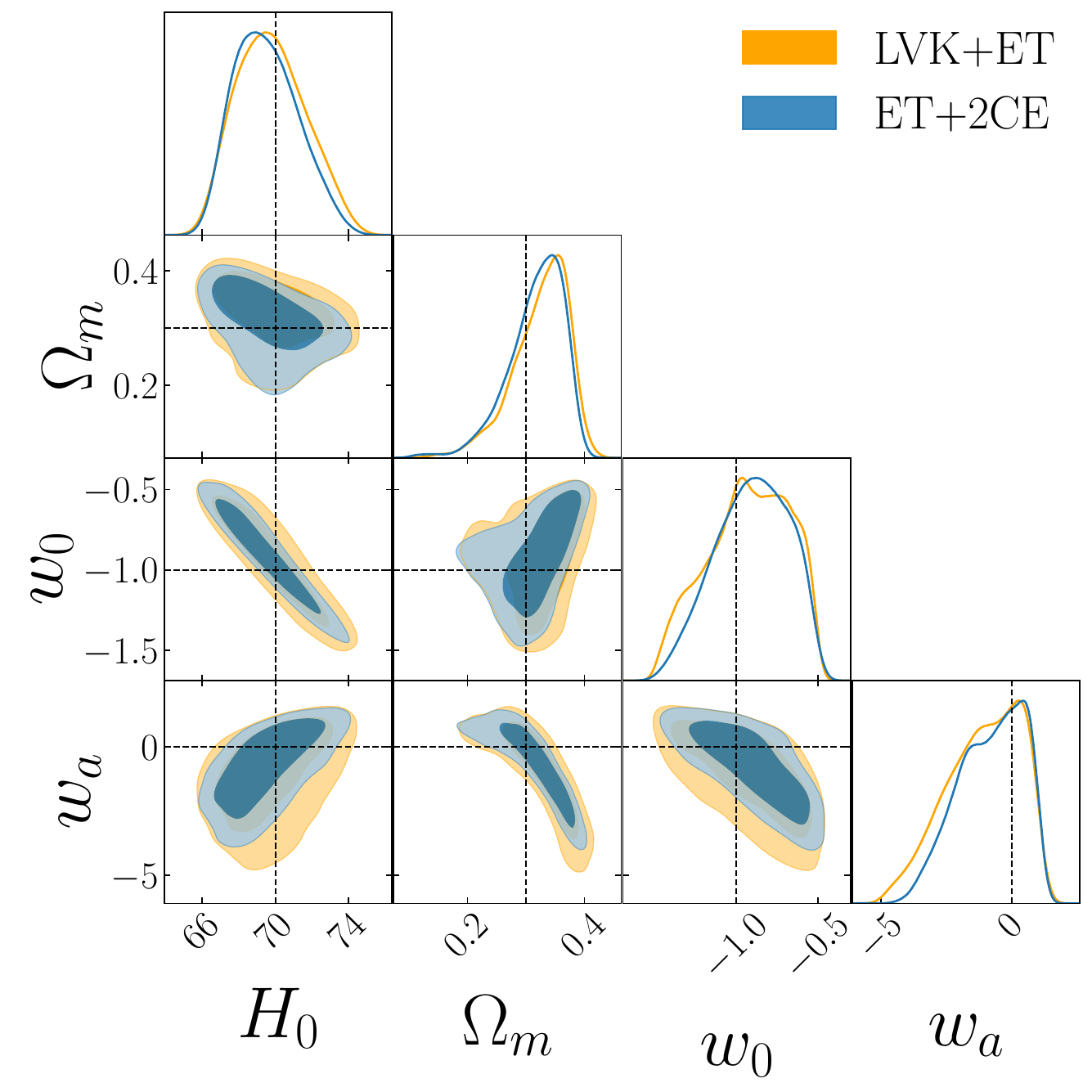}
    \caption{Posteriors for $w_0w_a$CDM, for LVK+ET and ET+2CE.
    Notice that one single third generation GW observatory increases the precision substantially, but the gain of adding further third generation detectors is more modest. See also Table~\ref{tab:results}.
    }
    \label{fig:et2ce2}
\end{figure}

The constraints under these configurations for $\Lambda$CDM are shown in Fig.~\ref{fig:et2ce}, and for $w_0w_a$CDM in Fig.~\ref{fig:et2ce2}. They
demonstrate our ability to recover the fiducial cosmology. In fact, for $\Lambda$CDM, in both detector cases we recover $H_0$ with sub-percent level precision, and $\Omega_m$ to 3--5\%.
Even though most events of LVK O5, given the expected sensitivities, should have a signal-to-noise ratio below 2, they contribute significantly towards triangulating the source positions, aiding in event localization.
A summary of the results for all considered networks is provided in Table~\ref{tab:results}.
For $w_0w_a$CDM in the LVK+ET (ET+2CE) network, we achieve an average precision of 2.8\% (2.5\%) in \(H_0\), of 13\% (14\%) in \(\Omega_m\) and of 30\% (24\%) in \( w_0 \), and can constrain \( w_a \) to be roughly within [$-3$,1] ([$-2$,1]).
Although the inclusion of dark energy parameters introduces additional uncertainties, the precision of the overall results remains significant, given the complexity of the extended model and the associated degeneracies.

\begin{table}%[H]
    \centering
    {\renewcommand{\arraystretch}{1.5}%
    \begin{tabular}{ccccc}
    \hline
     \rowcolor{lightgray}
     Network & $H_0\,[\hunit]$ & $\Omega_m$ & $w_0$ & $w_a$ \\ \hline
     LVK O5 (only $H_0$) & $70.4\pm3.0$ & --- & --- & --- \\
     LVK O5 (\(\Lambda\)CDM) & $68.3^{+4.2}_{-5.0}$ & $0.44^{+0.15}_{-0.37}$ & --- & --- \\
     LVK+ET (\(\Lambda\)CDM) & $69.96\pm0.67$ & $0.302^{+0.014}_{-0.016}$ & --- & --- \\
     ET+2CE (\(\Lambda\)CDM) & $69.99\pm0.42$ & $0.301\pm0.010$ & --- & --- \\
     LVK+ET (\(w_0w_a\)CDM) & $69.7^{+1.6}_{-2.3}$ & $0.325^{+0.058}_{-0.029}$ & $-0.92^{+0.35}_{-0.21}$ & $-1.06^{+2.00}_{-0.91}$ \\
     ET+2CE (\(w_0w_a\)CDM) & $69.5^{+1.5}_{-2.0}$ & $0.317^{+0.060}_{-0.029}$ & $-0.90\pm0.22$ & $-0.80^{+1.70}_{-0.81}$ \\
         \hline
    \end{tabular}
    }
    \caption{
    Final constraints on $H_0$, $\Omega_m$, $w_0$, and $w_a$ for the considered detector networks. We remark that LVK O5 alone is capable of a 4\% measurement of $H_0$, and that one single third generation GW observatory increases the precision by an order of magnitude in both $H_0$ and $\Omega_m$, and can make competitive measurements of both $w_0$ and $w_a$.}
    \label{tab:results}
\end{table}

\begin{figure}
    \centering
    \includegraphics[width=1.0\linewidth]{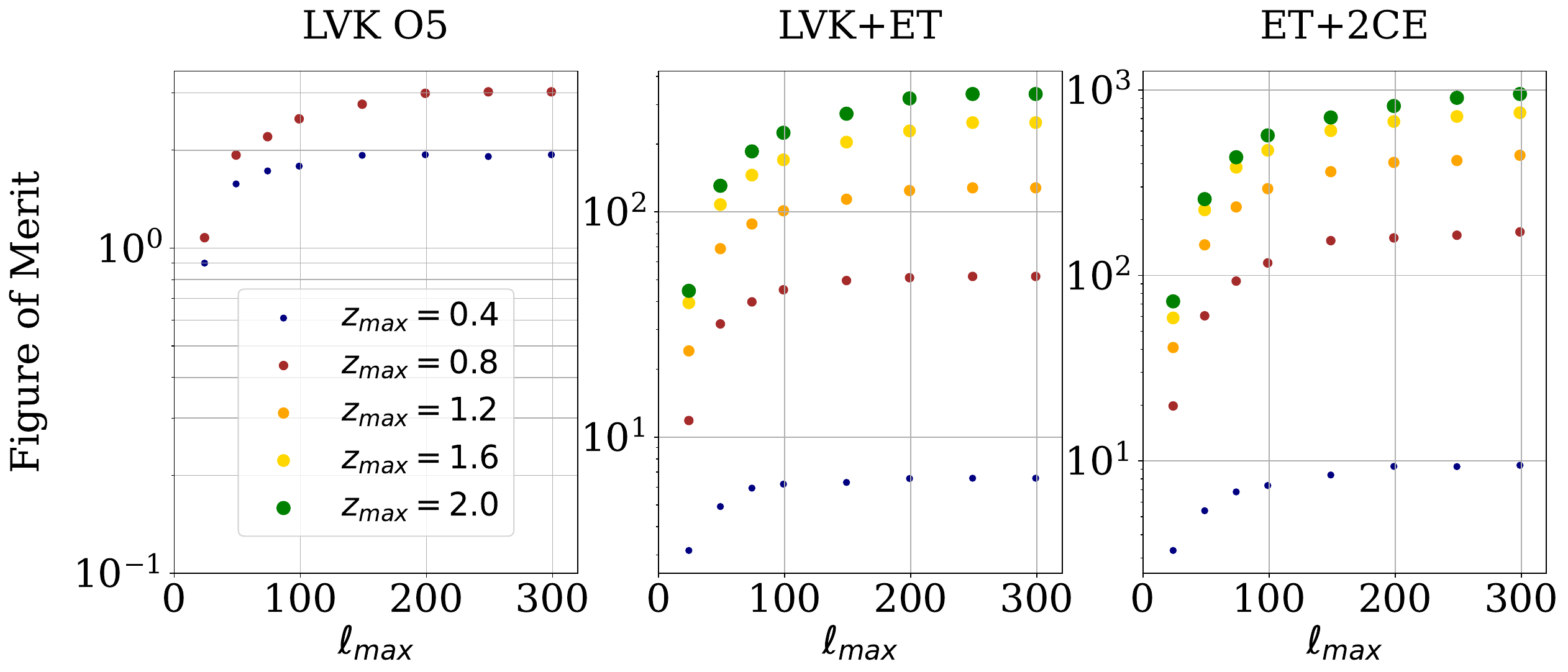}
    \caption{Figures of Merit for LVK (left), LVK+ET (center) and for the ET+2CE network (right), for different redshift and scale cuts.}
    \label{fig:et2ce_fom}
\end{figure}

We have also conducted our analyses with a set of different scale cuts, both in terms of the redshifts and bandpowers $\ell$. Fig.~\ref{fig:et2ce_fom} shows how the Figure-of-Merit,
\begin{equation}
    \label{eq:FoM}
    {\rm FoM} \equiv 1/\sqrt{|{\rm Cov}(H_0,\Omega_m)|} \; ,
\end{equation}
changes as we apply different redshift and scale cuts.
For LVK we include in our analysis modes up to $\ell=400$, but the left plot  of Fig.~\ref{fig:et2ce_fom} shows that the information saturates at around $\ell \sim 200$.
For the LVK+ET and  ET+2CE networks the analysis includes multipoles up to \( \ell \leq 700\), but the center and right plots of Fig.~\ref{fig:et2ce_fom} shows that the constraint converges around $\ell=300$ in both cases.
As for the redshift range, LVK already saturates for $z \lesssim 0.8$. However, the scenarios including ET may still benefit from galaxy or quasar catalogs that reach deeper than $z=2$ (which is the limit of the light cones we used in this work).

%%%%%%%%%%%%%%%%%%%%%%%%%%%%
\subsection{Consistency checks and possible systematics}
\label{sec:checks}

In this section, we examine the main potential sources of systematic errors and the effects of different scale cuts in our cross-correlation analysis. All consistency checks were conducted using the ET+2CE configuration, which is the case most sensitive to systematic uncertainties due to its higher precision.

\begin{figure}
    \centering
    \includegraphics[width=0.65\linewidth]{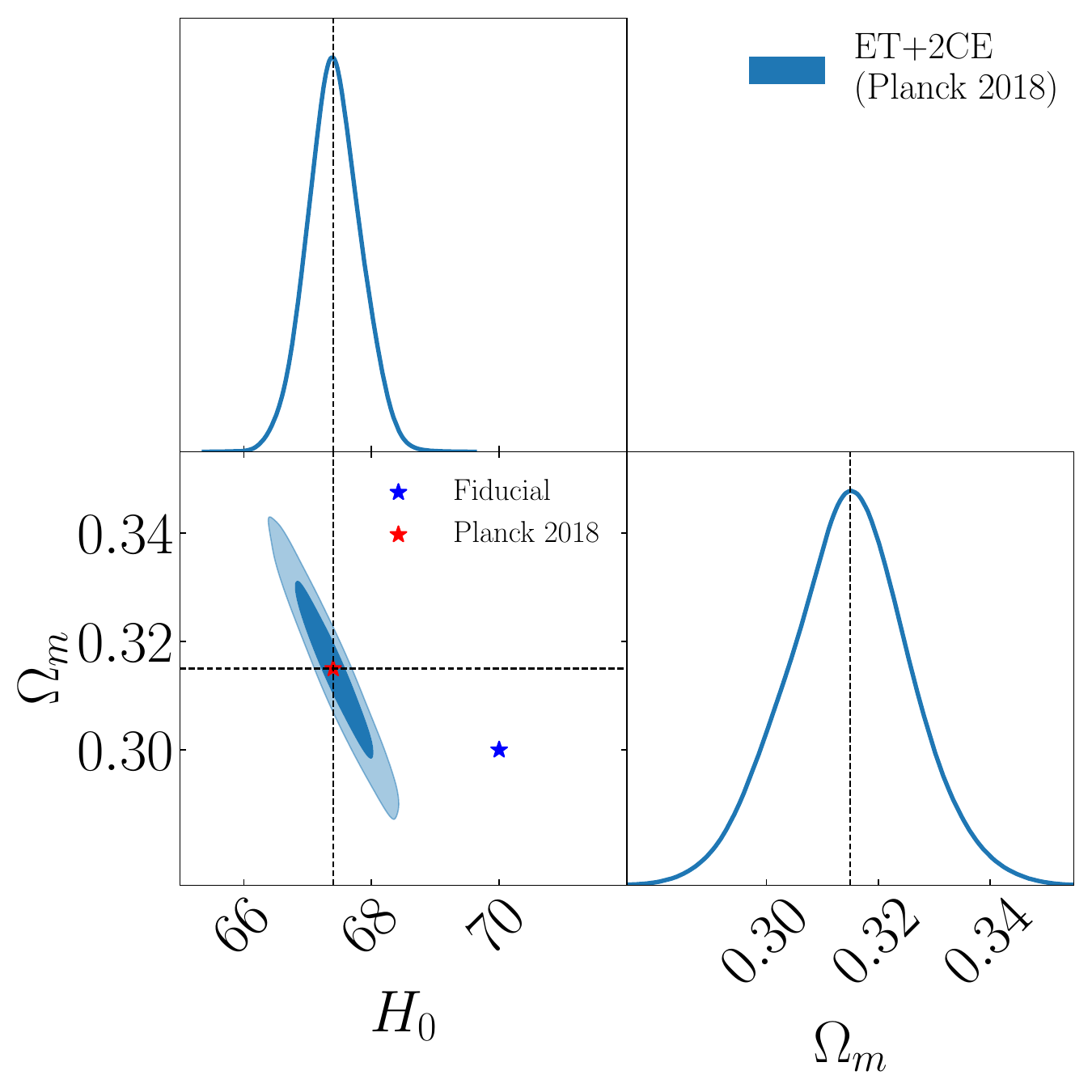}
    \caption{Recovering the cosmology from Planck 2018's best fit: even when using the cross-correlations from the fiducial cosmology \eqref{eq:fiducial_cosmology} as theoretical input (blue marker), we were able to recover the correct parameters for ET+2CE events simulated in a Planck 2018 cosmology (red marker). In blue, the final posteriors for $\Lambda$CDM.}
    \label{fig:et2ce_newcosmo}
\end{figure}

First, we emphasize that our analysis remains robust even if the fiducial cosmology used to generate the cosmic structures in the simulations differs from the true underlying cosmology.
To demonstrate this, we generated a pair of galaxy-BBH mock catalogs using Planck 2018's best-fit values $H_0=67.4\,\hunit$, and $\Omega_m=0.315$.
We then repeated our analysis using as data the cross-correlations obtained for the Planck mocks, $C_\ell^{gb}(\theta_{\rm Planck})$, and using as theory the set of $C_\ell^{gb}(\theta_{\rm fid})$ derived from simulations that used the fiducial cosmological parameters \eqref{eq:fiducial_cosmology}, where we used Eq.~\eqref{eq:alm_D4} to take into account the effect of the shifts in the redshift-distance relationships in the different cosmologies on the angular power spectra.
Fig.~\ref{fig:et2ce_newcosmo} shows that even when using a clustering model that is different than the real one, we are still able to recover the correct values of $H_0$ and $\Omega_m$, with negligible bias and with a precision similar to that of our main result: $H_0=67.41\pm0.42\,\hunit$, $\Omega_m=0.315\pm0.011$.
In other words: even with a  different shape for the power spectrum compared to the data, the peak of the galaxy-dark siren cross-correlation -- determined solely by the Hubble diagram
-- is sufficient to accurately determine the cosmological parameters. This is not only a test of the robustness of the Peak Sirens method, but also a validation of the procedure outlined in Section 2.2.

Another test we performed consists on removing all true host galaxies from our catalogs when computing the cross-correlations. We confirmed that this choice does not significantly impact the constraining power of our analysis, primarily due to the typically large solid angle errors associated with our BBH detections. For instance, as shown in Fig.~\ref{fig:bbh_catalogs}, approximately \(99\%\) of LVK O5 events exhibit angular uncertainties of \(\Delta\Omega > 1\,\,\text{deg}^2\).
At a redshift of \(z=1\), this corresponds to a circular region with a radius of about \(15\,\,\text{Mpc}\).
Consequently, the signal is not confined to scales corresponding to individual galaxies. Thus, our methodology demonstrates robustness against the challenges posed by catalog incompleteness, which could be a concern for various dark siren approaches~\cite{bera2020incompleteness,Schutz:1986gp,abbot2017,PhysRevD.101.122001,Gray_2022}.

In addition, we conducted two independent tests to evaluate potential sources of systematic errors in our analysis. While the data vector remains consistent with our primary analysis, the theoretical model is recalculated using the average of a new set of 1000 galaxy/BBH simulations generated for these specific scenarios. The results are then compared to those obtained from our standard configuration.

\begin{figure}
    \centering
    \includegraphics[width=0.49\linewidth]{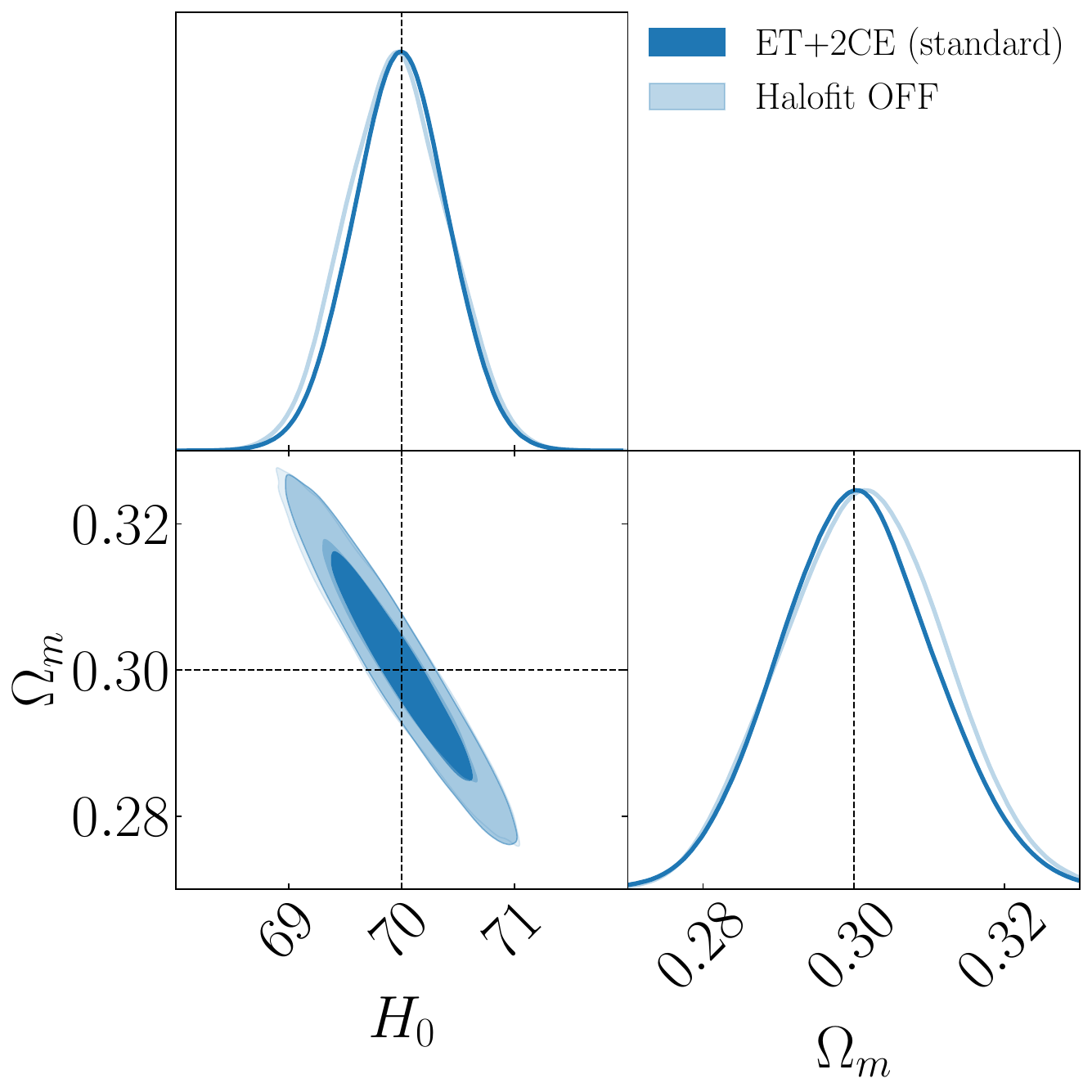}
    \includegraphics[width=0.49\linewidth]{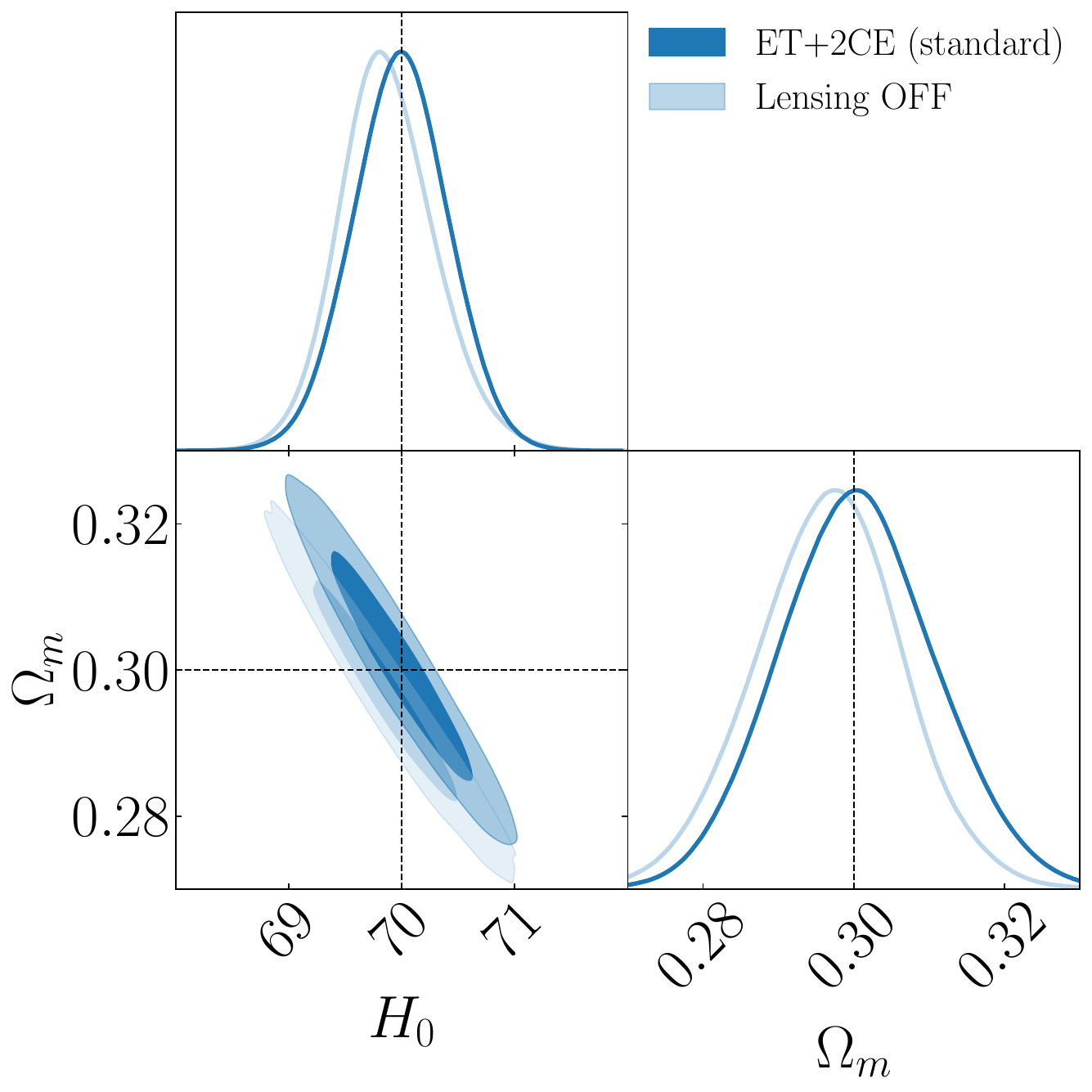}
    \caption{Possible source of biases, assuming ET+2CE. \emph{Left:} removing the \texttt{HALOFIT} corrections to the theoretical $C_\ell$ results in no bias. \emph{Right:} ignoring the BBH lensing magnification effect on the $C_\ell$. For this case, the biases are exaggerated by completely neglecting the lensing effect.
    In reality, the bias will be proportional to the uncertainties in the modeling of the magnification effect, which should have a much smaller impact than neglecting it completely.
    }
    \label{fig:halofit_lens_biases}
\end{figure}

The first test examines the impact of modifying the modeling of the matter power spectrum by disabling the small-scale corrections due to non-linear structure formation implemented by means of the \texttt{HALOFIT} model.
We simply switch off these non-linear corrections in the theoretical model (which are then computed from simulations that use a linear matter power spectrum), while using as data the simulations made with the non-linear (\texttt{HALOFIT}) spectrum.
From this analysis we have obtained the same results from our standard analysis (in which the non-linear corrections are included in the theoretical input) -- see the left plot of Fig.~\ref{fig:halofit_lens_biases}.
This highlights the insensitivity of the cross-correlation signal peak to non-linear corrections, demonstrating the robustness of the Peak Sirens method against different modeling of structure formation in the non-linear regime.

We also investigated the effects of not applying the weak lensing correction to the dark sirens $D_L$, as described in Section~\ref{sec:weaklensing}.
The constraints obtained for this exercise are $H_0 = 69.87^{+0.50}_{-0.49}\,\hunit$ and $\Omega_m = 0.3010\pm0.0098$ -- see the plot on the right of Fig.~\ref{fig:halofit_lens_biases}.
We remark that, for this test, the impact is exaggerated by completely neglecting the lensing effects.
In reality, the bias will be proportional to the uncertainties in the modeling of lensing, which should carry a much smaller impact than neglecting this effect completely.
For instance, different methodologies for computing the magnification PDF were compared in~\cite{Alfradique:2024fkb}, and while the skewness could change by $\sim 10\%$ depending on the approach, the position of the peak, which could affect our method, is almost unchanged.

Although these test indicate a remarkable robustness of the Peak Sirens method, as we increase the number of events and the precision of the dark siren data set, care must be taken in order to properly account for the contribution from such effects and to mitigate possible systematics.
We leave for future studies a more in-depth investigation of how much residual bias is left due to different models for non-linearities in the power spectrum and for the lensing PDF.

%\bigskip

%%%%%%%%%%%%%%%%%%%%%%%%%%%%%
\section{Discussion and conclusions}
\label{sec:conclusion}

We have shown how to harness the joint power of galaxy surveys and GW detections to constrain $H_0$ and other cosmological parameters.
By slicing the galaxy survey in redshift bins, and the BBH events in distance bins, we are able to compute the cross-correlations of both tracers in different radial bins.
When a redshift bin matches a distance bin, which happens only if the correct cosmological model is used to map one onto the other, the cross-correlation between galaxies and BBHs peaks -- a manifestation of the fact that the correlation function increases steeply at short distances, which is a universal feature of the large-scale structures.
We checked that the Peak Sirens method is robust against changes in the shape of the matter power spectrum due to different cosmologies: the key feature detected by this method depends chiefly on the fact that the correlation function increases sharply at small distances.

Our results show that, assuming $\Lambda$CDM and projected O5 sensitivity, we expect LVK should be able to measure $H_0$ with 4\% precision by using an external dataset to fix $\Omega_m$ or with 7\% precision by itself. This has the potential to provide a significant new outlook on the current $H_0$ tension.
Looking further into the future, the upcoming third generation GW detectors should provide exquisite constraints on some key cosmological parameters with the cross-correlation technique.
In particular, a single such detector operating in tandem with LVK can constrain $H_0$ to 1\% precision assuming $\Lambda$CDM, and to within 3\% assuming a very flexible $w_0w_a$CDM model with uninformative priors. It can also constrain $\Omega_m$, $w_0$ and $w_a$ with very competitive precision.

As for the next steps, we intend to further assess the impact of various physical effects that could impact the cross-correlations, and to use this standard ruler for measuring the Hubble constant with real data. These subdominant effects include the role of the assumed redshift distribution of BBHs, as well as the BBH and galaxy biases. A relative BBH/galaxy bias could arise, for example, if there is a preference for BBH hosts that depends on luminosity, and could be included in simulations where these types of galaxy properties are also modeled. In this paper we assume all galaxies are equally likely to host BBHs, but this assumption should be dropped in future detailed studies on the impact of the BBH bias.

The likely independence on population model, or on the identification of the true host galaxy, are some of the features that distinguish the Peak Sirens method from the line-of-sight prior method. While the latter relies for the most part on coincidences between the position of objects parallel to the line-of-sight, the former takes advantage of both parallel and transverse correlations, and of how these drop quickly with relative distance. While the two methods can be complementary, we expect different limitations to play a role, and each one could be more efficient in different regimes, depending on factors such as the incompleteness of the galaxy catalog or the number of GW events detected with sufficient accuracy.
We leave a full comparison between these methods for future work.

Another interesting venue for investigation with the cross-correlation method is to study how well we could detect large-scale deviations from General Relativity. The main imprint of scalar-tensor theories in tensor modes are a modified friction in the GW propagation across large distances, which would make the distance one would attribute to the GW source from its measured amplitude to differ from the luminosity distance. In particular, the parameter $\Xi = D_{\rm GW}/D_L$ is expected to be directly measured with third-generation detectors from observations of bright standard sirens when combined with other distance indicators \cite{Matos:2023jkn}, or with gravitational slip measurements \cite{Matos:2022uew} to less than percent level. With the forthcoming $\mathcal{O}(10^6)$ observations of supernovae, an interesting extension of these ideas would be to study the cross-correlation of BBHs and those standard candles for measuring $\Xi$.
%%%%%%%%%%%%%%%%%%%%%%%%%%%%%

\section*{Acknowledgments}

We would like to thank Arthur Loureiro for extensive help with GLASS and for many helpful comments in our early draft. We also thank Simone Mastrogiovanni for several useful suggestions.
JF and IT would like to thank CNPq for financial support.
RA and RS are supported by grants from FAPESP and CNPq.
ISM also acknowledges support from FAPESP. MQ is supported by the Brazilian research agencies Fundação Carlos Chagas Filho de Amparo à Pesquisa do Estado do Rio de Janeiro (FAPERJ) project E-26/201.237/2022 and CNPq.
This research was supported by resources supplied by the Center for Scientific Computing (NCC/GridUNESP) of the São Paulo State University (UNESP).

\bibliographystyle{JHEP}
\bibliography{stuff.bib}

\FloatBarrier
\appendix
\section{Appendix: Cross-correlation signal} \label{sec:appendix}

We use the sample covariance derived from simulations to fit data generated through the same process, hence the covariance encapsulates the noise characteristics inherent to the simulated data. This can lead to artificially tight constraints if we continue increasing \(\ell_{\text{max}}\) in the likelihood, despite the fact that small angular scales are not a reliable source of information due to significant uncertainties in the BBH positions.
Therefore, it is essential to identify the scales that provide meaningful information. Starting from the general multi-tracer angular power spectrum in equation \eqref{eqn:angpow}, we can express the galaxy-BBH angular power spectrum as:
\begin{equation}
    \label{eqn}
    C_\ell^{gb} (z_g,z_b)=\Gamma_\ell^{ij} (z_g,z_b)-\delta_{gb}\,\delta_{z_g,z_b}\,\bar{N}^g(z_g) \; ,
\end{equation}
where \(z_g\) and \(z_b\) are the 100 thin redshift bins defined in \eqref{eq:zbins}. In this expression, we assume that the tracers \(i\) and \(j\) are different, meaning that the shot noise associated with the cross-correlation between them is negligible. However, in our case, the BBH events are selected from the galaxy mocks, implying that both tracers share the same bias parameter. Consequently, there may be a non-trivial shot noise contributing to the cross-correlation, as per \eqref{eqn:cros_corr}.

A final remark with respect to the signal we measure is regarding masking effects. As already pointed out in Section~\ref{sec:gal_bbh_mocks}, we apply a mask on the galaxy catalogs corresponding to approximately one-third of the sky. This implies two things: the first is that we are also roughly discarding $1/3$ of the BBH events which fall on the mask area, since there are no galaxies in this region to correlate these BBHs with. The second is that what we actually measure now is a \textit{pseudo-}$C_\ell^{ij}$, which has artificial signals due to the masking on the galaxy catalogs. The pseudo-$C_\ell^{ij}$, denoted by $\tilde C_\ell^{ij}$, relates to the \textit{true} underlying $C_\ell^{ij}$ by~\cite{Alonso_2019}:
\begin{equation}
\label{eq:pseudo-cl}
    \tilde C_\ell^{ij}=\frac{1}{(2\ell+1)}\sum_{\ell'}\,M_{\ell\,\ell'}^{ij}\,C^{ij}_\ell\,\,.
\end{equation}
Therefore, the pseudo-$C_\ell^{ij}$ causes a mix on the multipoles, given by the \textit{mode-coupling matrix} $M_{\ell\,\ell'}^{ij}$ defined by:
\begin{equation}
\label{eq:mode_coupling}
    M_{\ell\,\ell'}^{ij}=\frac{2\ell'+1}{4\pi}\sum_{\ell''}(2\ell''+1)\,W^{ij}_{\ell''}\begin{pmatrix}
        \ell & \ell' & \ell'' \\
        0 & 0 & 0
    \end{pmatrix}^2\,\,,
\end{equation}
where $W^{ij}_{\ell''}$ is the angular power spectrum of the mask, and the matrix denotes the Wigner-3j symbols. In order to obtain the \textit{true} $C_\ell$, one must invert the relation~\eqref{eq:pseudo-cl} -- in other words, invert the mode-coupling matrix~\eqref{eq:mode_coupling}. Furthermore, we need to deal with the fact that different multipoles are correlated to each other, breaking the validity of~\eqref{eqn:MultiTracerChi2}. To mitigate these effects and obtain an accurate estimate of the true power spectra, we employ the \texttt{NaMaster} code\footnote{\url{https://github.com/LSSTDESC/NaMaster}} \cite{Alonso_2019}, which corrects for the mode mixing caused by the mask.  We verified that bandpowers of width 5 provide sufficient precision. Furthermore, the mask induces distortions at low multipoles, and thus we restrict the likelihood analysis to \(\ell \geq 7\), effectively excluding the first bandpower.

\end{document}